\documentclass[onecolumn,superscriptaddress,prd,aps,floats,floatfix,nofootinbib,amsmath,amssymb]{revtex4}
\usepackage{graphicx}
\usepackage{color}
\usepackage{amsmath,amssymb}
\usepackage{array}

\newcommand{\vev}[1]{\left\langle #1 \right\rangle}

\newcommand{\TeV}{\text{TeV}}
\newcommand{\GeV}{\text{GeV}}

\newcommand{\crit}{{\rm crit}}
\newcommand{\tr}{\mbox{tr}}
\newcommand{\gtwo}{I\kern-.1em I\,}

\newcommand{\beq}{\begin{eqnarray}}
\newcommand{\eeq}{\end{eqnarray}}
\newcommand{\bpm}{\begin{pmatrix}}
\newcommand{\epm}{\end{pmatrix}}
\newcommand{\cl}{\, \rm C.L.}

\begin{document}

\title{Top-Mode Pseudos at the LHC}
\author{Hidenori S. Fukano}
\thanks{\tt fukano@kmi.nagoya-u.ac.jp}
      \affiliation{ Kobayashi-Maskawa Institute for the Origin of Particles and 
the Universe (KMI) \\ 
 Nagoya University, Nagoya 464-8602, Japan.}
\author{Shinya Matsuzaki}\thanks{\tt synya@hken.phys.nagoya-u.ac.jp}
      \affiliation{ Institute for Advanced Research, Nagoya University, Nagoya 464-8602, Japan.}
      \affiliation{ Department of Physics, Nagoya University, Nagoya 464-8602, Japan.}

\begin{abstract}
We discuss LHC phenomenologies of the Top-Mode Pseudos $(h^0_t,A^0_t)$, 
composite pseudo Nambu--Goldstone bosons 
predicted in the model recently proposed 
in a framework of the top quark condensation. 
The CP-even Top-Mode Pseudo, dubbed tHiggs ($h^0_t$), is identified with 
the 126 GeV Higgs boson at the LHC.  
We analyze the coupling properties of the tHiggs 
in comparison with the currently available data reported from LHC Run-I 
to find that the tHiggs can be consistent with the LHC Higgs boson. 
The mass formula 
relating masses of the Top-Mode Pseudos 
allows us to place 
an indirect limit on the mass of the CP-odd Top-Mode Pseudo ($A^0_t$) from 
the constraint on the tHiggs coupling strengths.  
The presence of the mass formula also significantly affects the coupling property of $A^0_t$, 
which ensures $A^0_t$ weakly coupled to the Standard Model particles. 
The direct limit on the mass of $A^0_t$ is placed by
data on searches for new resonances in several channels at the LHC Run-I. 
We find the lower mass bound from both 
the indirect and direct limits, $m_{A^0_t} \gtrsim 560\,\GeV$.  
The discovery channel of $A^0_t$ in the upcoming LHC Run-II is also addressed. 
\end{abstract}

\maketitle

\section{Introduction}

A 126 GeV Higgs boson has been discovered at the LHC (so-called LHC Run-I)
by the ATLAS~\cite{Aad:2012tfa} and CMS experiments~\cite{Chatrchyan:2012ufa}. 
It has so far been reported that the LHC Higgs boson has the properties 
compatible with the Higgs boson in the Standard Model (SM). 
After the discovery of the 126 GeV Higgs boson, 
one of the primary targets for future collider experiments such as the upcoming next LHC run (LHC Run-II)  
is to reveal the dynamical origin of the Higgs boson 
responsible for the mass generation of the SM particles 
and 
discover the related new particles beyond the SM. 

One key hint at accessing such a dynamical origin of the Higgs boson 
would be deduced from an observed coincidence among scales of top quark, Higgs boson,  
and weak gauge boson masses; 
of the SM particles, they are the only ones roughly on the same order. 
Top quark condensation~\cite{Miransky:1988xi,Miransky:1989ds,Nambu:1989jt,
Marciano:1989xd,Marciano:1989mj,Bardeen:1989ds} naturally provides 
such a close relation between those mass scales.  
The top quark condensate model was proposed~\cite{Miransky:1988xi,Miransky:1989ds}
to predict the top quark mass to be on the order of the electroweak symmetry breaking scale, 
which was before the discovery of the top quark. 
However, 
the original top quark condensate model is somewhat far from a realistic situation; 
the predicted value of the top quark mass is too large compared to the experimental value.  
In addition, a Higgs boson predicted as a $t \bar{t}$ bound state 
has the mass in a range of $m_t < m_H < 2 m_t $, 
which cannot be identified with the 126 GeV Higgs boson at the LHC. 

Recently,
a new class of the top quark condensate model was 
proposed~\cite{Fukano:2013aea}~\footnote{At almost the same time as \cite{Fukano:2013aea},
a similar model was proposed in a slightly different context~\cite{Cheng:2013qwa}.
}, 
where the realistic top quark mass is obtained by  
the top-seesaw mechanism~\cite{Dobrescu:1997nm,Chivukula:1998wd}
and 
a composite Higgs boson emerges as a pseudo Nambu--Goldstone boson (PNGB)
associated with the spontaneous breaking of a global symmetry, 
which can be as light as the 126 GeV Higgs boson at the LHC. 
The model in Ref.\cite{Fukano:2013aea} is constructed from 
the top and bottom quarks $q=(t,b)$ and a (vectorlike) $\chi$-quark, 
a flavor partner of the top quark having 
the same SM charges as those of the right-handed top quark, 
which form a four-fermion interaction,
$G_{4f} (\bar{\psi}^i_L \chi_R)(\bar{\chi}_R \psi^i_L)$, 
where 
$\psi^i_L \equiv (t_L , b_L, \chi_L)^{T\,i}\,\,( i = 1,2,3)$. 
The model possesses a global symmetry $U(3)_{L} \times U(1)_{\chi_R}$, 
which is spontaneously broken down to $U(2)_{L} \times U(1)_{L + \chi_R}$ 
by the quark condensations 
$\vev{\bar{\chi}_R t_L} \neq 0$, and $\vev{\bar{\chi}_R \chi_L} \neq 0$, 
triggered by the supercritical four-fermion coupling $G_{4f} > G_\crit$, 
where $G_\crit$ is the critical coupling. 
The associated five Nambu--Goldstone bosons (NGBs) emerge as bound states of the quarks,  
in addition to a composite heavy Higgs boson corresponding to the $\sigma$ mode of 
the usual Nambu-Jona-Lasinio (NJL) model~\cite{Nambu:1961tp}. 

Three of those NGBs are eaten by the electroweak gauge bosons 
when the subgroup of the symmetry is gauged by the electroweak symmetry, 
while two of them become PNGBs and remain as physical states, dubbed ``Top-Mode Pseudos". 
Those two Top-Mode Pseudos acquire their masses due to additional terms 
that explicitly break the $U(3)_{L} \times U(1)_{\chi_R}$ symmetry 
in such a way that the vacuum aligns to break the electroweak symmetry 
by $\vev{\bar{\chi}_R t_L} \neq 0$. 
One of them is a CP-even scalar, ``tHiggs" ($h^0_t$), 
which is identified as the 126 GeV Higgs boson at the LHC,  
while the other one is a CP-odd scalar ($A^0_t$) having the same quantum number 
as that of other CP-odd scalars as 
in the minimal supersymmetric standard model (MSSM) and the two-Higgs doublet model (2HDM). 
These two Top-Mode Pseudos are expected to be the low-lying spectra 
and 
the CP-odd scalar $A^0_t$ will be a new particle, 
which can be the phenomenological consequence for the model 
to be tested at the LHC. 

The model in Ref.\cite{Fukano:2013aea} predicts a notable relation 
between masses of two Top-Mode Pseudos at the tree-level: 
\beq
m^{(0)}_{h^0_t} = m^{(0)}_{A^0_t} \sin \theta
\,,\nonumber 
\eeq
where the angle $\theta$ is a model parameter related to the presence of the condensate, 
$\vev{\bar{\chi}_R q_L} \neq 0$, which causes the electroweak symmetry breaking.  
The above mass formula is established at the tree level of the perturbation with 
respect to couplings explicitly breaking the $U(3)_{L} \times U(1)_{\chi_R}$ symmetry.   
The next-to leading order corrections, 
especially coming from terms introduced to generate the top quark mass, 
gives rise to significant corrections to the $h^0_t$ mass, 
so that the tree-level $h^0_t$ mass is set 
so as to realize the mass at around 126 GeV at the one-loop level, 
say, $m^{(0)}_{h^0_t} \simeq 230\,\GeV$~\cite{Fukano:2013aea} 
[see also Eq.(\ref{hmass:tree})].  
On the other hand,  the $A^0_t$ mass does not get large corrections~\cite{Fukano:2013aea}, 
and hence the one-loop mass is almost the same as the tree-level one, 
$m_{A^0_t} \simeq m^{(0)}_{A^0_t}$   
[see Eq.(\ref{Amass:tree})]. 

The angle $\theta$ in the above formula  
also controls the size of deviation of the tHiggs couplings 
from the SM Higgs ones~\cite{Fukano:2013aea}:
\beq
\frac{g_{h^0_tVV}}{g_{h_{_{\rm SM}}VV}} 
= 
\frac{g_{h^0_t bb}}{g_{{h_{\rm SM}} bb}} 
= 
\frac{g_{h^0_t \tau\tau}}{ g_{h_{_{\rm SM}} \tau\tau}}  
= 
\cos \theta
\quad , \quad
\frac{g_{h^0_t tt}}{g_{h_{{\rm SM}} tt}}  
= 
\frac{2\cos^2\theta -1}{\cos\theta} 
\,. \nonumber
\eeq 
Note that the tHiggs couplings to the SM particles 
coincide with those of the SM Higgs boson 
in an extreme limit $\sin \theta \to 0$. 
In this limit, 
one can see from the above mass formula 
that $m^{(0)}_{A^0_t}/m^{(0)}_{h^0_t} \to \infty$, 
implying decoupling of the $A^0_t$ from the theory. 
Thus precise measurements of deviation from the SM Higgs couplings 
would be a crucial key for the presence of $A^0_t$ 
having the mass within the reach of the LHC search 
and 
would also place an indirect bound on the $A^0_t$ mass, 
in addition to the limits from the direct searches for $A^0_t$ at the LHC. 

In this paper, 
we discuss LHC phenomenologies of the Top-Mode Pseudos $(h^0_t,A^0_t)$ 
predicted in the model of Ref.\cite{Fukano:2013aea}.  
Identifying the CP-even Top-Mode Pseudo (tHiggs, $h^0_t$) 
with the 126 GeV Higgs boson at the LHC,  
we analyze the coupling properties of the tHiggs and compare them with 
the currently available data on the Higgs coupling measurements reported from LHC Run-I%
~\cite{ATLAS-CONF-2013-012,Aad:2013wqa,
ATLAS-CONF-2013-030,ATLAS2013108,TheATLAScollaboration:2013lia,
CMS-PAS-HIG-13-001,CMS-PAS-HIG-13-002,Chatrchyan:2013iaa,CMS:2013yea,
CMS-PAS-HIG-13-004,Chatrchyan:2013zna}. 
We evaluate the constraint on the tHiggs couplings to the SM particles 
to find that the tHiggs can be consistent with the LHC Higgs boson, 
allowing a size of the deviation (controlled by $\cos\theta$) from the SM Higgs couplings. 
An indirect limit on the mass of CP-odd Top-Mode Pseudo ($A^0_t$) 
is set through the mass formula 
($\sin\theta= m^{(0)}_{h^0_t}/m^{(0)}_{A^0_t}\simeq m^{(0)}_{h^0_t}/m_{A^0_t}$)  
to be $m_{A^0_t} \geq 563 \,\GeV$. 

We then calculate the production cross sections and partial decay widths of $A^0_t$ 
relevant to the LHC study 
and 
find that $A^0_t$ is dominantly produced via the gluon-gluon fusion process 
and mainly decays to $gg$ and $Z h^0_t$ for a low-mass range, 
$563 \,\GeV \leq m_{A^0_t} \leq 1\,\TeV$, 
and $t\bar{t},gg$ for a high mass range, $m_{A^0_t} > 1\,\TeV$. 
The total width of $A^0_t$ is shown to be quite smaller than 
that of the SM Higgs boson 
and other CP-odd scalars like $A^0$ as in the MSSM/2HDM. 
This is due to the presence of the mass formula displayed above, 
which is intrinsic to the Top-Mode Pseudos; 
the mass formula ensures $A^0_t$ weakly coupled to the SM particles. 
Furthermore, 
we place a direct mass bound on $A^0_t$ from 
the LHC Run-I data on direct searches for new resonances in several channels. 
We then find all the direct limits are milder than the indirect limit. 

The discovery channel of $A^0_t$ in the upcoming LHC Run-II is also addressed; 
a heavy $A^0_t$ with the mass $m_{A^0_t} \geq 1\,\TeV$ can be seen as a quite narrow resonance 
in the $A^0_t \to t \bar{t}$ or $A^0_t \to gg$ channel, 
while a light $A^0_t$ with the mass in a range $563 \,\GeV \leq m_{A^0_t} \leq 1\,\TeV$ 
may be measured via $A^0_t \to Z h^0_t$, 
which would be earlier than the discovery of other CP-odd scalars. 

This paper is organized as follows.
In Sec.~\ref{EFT-TMP}, 
we give a brief review of a phenomenological Lagrangian 
describing the Top-Mode Pseudos $(h^0_t, A^0_t)$  
based on the 
$[U(3)_{L} \times U(1)_{\chi_R}]/[U(2)_{L} \times U(1)_{L + \chi_R}]$ 
nonlinear sigma model~\cite{Fukano:2013aea}.  
In Sec.~\ref{Ph-ht0}, 
the coupling properties of $h^0_t$ are discussed 
in comparison with the currently available data from the Higgs coupling measurements at the LHC Run-I,   
and 
then we convert the result into the constraint on the mass of $A^0_t$ 
through the mass formula mentioned above. 
In Sec.~\ref{Ph-At0},
we compute the partial decay widths and production cross sections of $A^0_t$ 
relevant to the LHC study. 
The limits on the $A^0_t$ mass are then placed from the LHC Run-I data 
on direct searches for new resonances 
in several channels currently reported by the ATLAS and CMS experiments. 
The discovery channel of $A^0_t$ is also addressed in light of the upcoming LHC Run-II. 
Section~\ref{summary} is devoted to the summary of this paper.

\section{Phenomenological Lagrangian for the Top-Mode Pseudos}
\label{EFT-TMP}

In this section, 
we review a low-energy effective Lagrangian relevant to 
studying the LHC phenomenologies of the Top-Mode Pseudos~\cite{Fukano:2013aea}. 
The model proposed in Ref.\cite{Fukano:2013aea} consists of 
the top and bottom quarks $q^3=(t,b)$ and a (vectorlike) $\chi$-quark 
which is a flavor partner of the top quark having the same SM charges 
as those of the right-handed top quark.  
In addition to the kinetic term of those quarks, 
the model includes 
a four-fermion interaction $G_{4f} (\bar{\psi}^i_L \chi_R)(\bar{\chi}_R \psi^i_L)$, 
where $\psi^i_L \equiv (q^3_L, \chi_L)^{T\, i} =((t_L , b_L), \chi_L)^{T\,i}\,( i = 1,2,3)$.
The model then possesses a global symmetry $U(3)_{L} \times U(1)_{\chi_R}$ 
that is spontaneously broken down to $U(2)_{L} \times U(1)_{L+\chi_R}$ 
by the quark condensations $\vev{\bar{\chi}_R t_L} \neq 0$ and $\vev{\bar{\chi}_R \chi_L} \neq 0$, 
triggered by the supercritical four-fermion coupling $G_{4f} > G_\crit$ 
where $G_\crit$ is the critical coupling. 

As discussed in Ref.\cite{Fukano:2013aea}, the structure of the symmetry breaking pattern of 
the model is actually twisted; the gap equation derived from the four-fermion dynamics 
has a rotational invariance with respect to the fermion dynamical masses associated with 
the condensates $\vev{\bar{\chi}_R t_L} \neq 0$ and $\vev{\bar{\chi}_R \chi_L} \neq 0$, 
which corresponds to changing the basis of the left-handed quarks from the electroweak gauge base 
to a flavor base as $\tilde{\psi}_L = R(\theta) \cdot \psi_L$ 
by the orthogonal rotation of $R(\theta)$. 
The symmetry-breaking pattern thus looks like 
$U(3)_{\tilde{\psi}_L} \times U(1)_{\chi_R} \to U(2)_{\tilde{\psi}_L} \times U(1)_{\tilde{\psi}_L + \chi_R}$.   
The associated five NGBs emerge as bound states of the quarks,  
in addition to a composite heavy Higgs boson ($H^0_t$)
corresponding to the $\sigma$ mode of the usual NJL model. 
Three of the NGBs are eaten by the $W$ and $Z$ bosons 
when the electroweak gauge is turnd on, 
while the other two become massive due to some explicit breaking effects, 
which we call 
the Top-Mode Pseudos. 

Below the heavy composite Higgs mass scale (of ${\cal O}(1)\,\TeV$~\cite{Fukano:2013aea}), 
the model can be described by a nonlinear sigma model 
based on the coset space 
$G/H = [U(3)_{\tilde{\psi}_L} \times U(1)_{\chi_R}]/[U(2)_{\tilde{\psi}_L} \times U(1)_{\tilde{\psi}_L + \chi_R}] $.  
The representatives of the $G/H$ 
parametrized by NGB fields $\pi^a_t$ $(a=4,5,6,7,A)$ are
\beq
\xi_L = 
\exp\left[ 
- \frac{i}{f} \left( 
\sum_{a=4,5,6,7} \pi^a_t \lambda^a  + \frac{\pi^A_t}{2\sqrt{2}} \lambda^A 
\right) 
\right]
\quad , \quad
\xi_R = 
\exp\left[ 
\frac{i}{f} 
  \frac{\pi^A_t}{2\sqrt{2}} \lambda^A 
\right]
\,,\nonumber
\eeq
where $f$ is the decay constant associated with the spontaneous breaking $G/H$, 
$\lambda^a$ denotes the Gell-Mann matrices and 
\beq 
\lambda^A
= 
\bpm 0 & 0 & 0 \\ 0 & 0 & 0 \\ 0 & 0 & \sqrt{2} \epm
\,.\nonumber
\eeq
It is convenient to further introduce the ``chiral" field $U$ as
\beq 
U = \xi_L^\dag \cdot \Sigma \cdot \xi_R 
\quad
\text{with \,\, $\Sigma = \frac{1}{\sqrt{2}} \lambda^A$}
\,.\label{def-NLsM-U}
\eeq
The transformation properties of $\xi_{L,R}$ and $U$ under $G$ are given by
\beq
\xi_L \to  h(\pi_t,\tilde{g}) \cdot\xi_L \cdot g_{\tilde{3L}}^\dag
\quad , \quad
\xi_R \to  h(\pi_t,\tilde{g}) \cdot\xi_R \cdot g_{1R}^\dag
\quad , \quad
U \to g_{\tilde{3L}} \cdot U \cdot g^\dagger_{1R}
\,,\label{transform-NLsM-U}
\eeq
where 
$\tilde{g}=\{ g_{\tilde{3L}},g_{1R}\}\,,\,g_{\tilde{3L}} \in U(3)_{\tilde{\psi}_L} \,,\,g_{1R} \in U(1)_{\chi_R}$,  
and $h(\pi_t,\tilde{g}) \in H$. 
The $G$-invariant Lagrangian is thus constructed 
in terms of the NGBs to the lowest order of derivatives as 
\beq 
\frac{f^2}{2} \tr\left[  \partial_\mu U^\dagger  \partial^\mu U\right]
\,.
\label{NLsM-Lag-Op2-0}
\eeq

When the electroweak symmetry is turned on, 
the covariant derivative acting on $U$ is given by 
\beq
D_\mu U
\equiv
R(\theta) \left[
\partial_\mu 
- i g \sum^3_{a=1} W^a_\mu 
\left(
\begin{array}{cc|c}
&&0\\
\mbox{\raisebox{1.5ex}{\smash{\large$\mspace{5mu}\tau^a/2$}}}&&0\\ \hline
\mspace{-20mu}0&\mspace{-20mu}0&0
\end{array}
\right)
+ i g' B_\mu \bpm 1/2 & 0 & 0 \\ 0 & 1/2 & 0 \\ 0 & 0 & 0 \epm 
\right]
R^T(\theta)
\cdot
U 
\,,\quad 
R(\theta) =
\bpm
\cos \theta & 0 & - \sin \theta \\[1ex]
0 & 1 & 0 \\[1ex]
\sin \theta & 0 & \cos \theta 
\epm
\,,\label{covariant-derivative-U}
\eeq 
where $W_\mu$ and $B_\mu$ are the $SU(2)_L$ and $U(1)_Y$ gauge boson fields 
with the gauge couplings $g$ and $g'$, respectively. 
Thus, the Lagrangian Eq.(\ref{NLsM-Lag-Op2-0}) 
is changed to the covariant form 
\beq
{\cal L}_{\text{NL$\sigma$M}}
=
\frac{f^2}{2} \tr\left[  D_\mu  U^\dagger  D^\mu  U\right]
\,.\label{NLsM-Lag-Op2}
\eeq
The NGBs $(z^0_t,w^\pm_t) \equiv (\pi^4_t \cos\theta + \pi^A_t \sin\theta\,,  
(\pi^6_t \mp i \pi^7_t )/\sqrt{2})$ are then eaten by the $Z$ and $W$ bosons, 
leading to the $Z$ and $W$ masses, 
\beq
m^2_Z = \frac{1}{4}(g^2 + g'^2) f^2 \sin^2\theta
\quad , \quad 
m^2_W =  \frac{1}{4}g^2 f^2 \sin^2\theta
\,.\nonumber
\eeq
These mass formulas imply  
\beq
v^2_{_{\rm EW}} 
= f^2\sin^2\theta 
\,,\label{def-vEW}
\eeq
which is set by the Fermi constant $G_F$ as 
$v_{_{\rm EW}}=(\sqrt{2}G_F)^{-1/2} \simeq 246\,\GeV$~\cite{Beringer:1900zz}.
As seen from Eq.(\ref{def-vEW}), 
the nonzero angle $\theta$ $(\sin\theta)$, 
rotating left-handed fermions in the gauge $(\psi_L)$ and flavor ($\tilde{\psi}_L$)  bases,  
dictates the electroweak symmetry breaking $(v_{_{\rm EW}})$ 
and hence is related to the vacuum alignment problem~\footnote{ 
As was noted in Ref.\cite{Fukano:2013aea}, 
the criticality $G_{4f}>G_\crit$ implies the $R$-rotational invariant condensation, 
$\vev{\bar{\chi}_R t_L}^2+ \vev{\bar{\chi}_R \chi_L}^2 \neq 0$, 
but not necessarily $\vev{\bar{\chi}_R t_L} \neq 0$, 
which is responsible for the electroweak symmetry breaking.  
The electroweak gauge interaction itself can contribute to lifting the degeneracy 
between the vacuum with $\vev{\bar{\chi}_R t_L} = 0$ 
and that with $\vev{\bar{\chi}_R t_L} \neq 0$, 
which 
in principle would require some extreme fine-tuning of the critical coupling as well as the angle $\theta$ 
and some explicit breaking parameters ($G'$ and $G''$ as introduced in Ref.\cite{Fukano:2013aea}). 
The explicit analysis on the vacuum alignment will be pursed in another publication.  
}. 

The remaining two NGBs $(h^0_t,A^0_t) \equiv (\pi^5\,,-\pi^4_t\sin\theta + \pi^A_t \cos\theta)$ 
will become PNGBs (Top-Mode Pseudos), 
through explicit breaking terms introduced appropriately to 
the underlying four-fermion dynamics~\cite{Fukano:2013aea}:
\beq 
\Delta {\cal L}_{\text{NL$\sigma$M}}
= 
f^2 
 \tr
\left[
c_1 (R^T U)^\dagger \chi_1 (R^T U)
+ 
c_2 \left( \chi^\dagger_2 (R^T U) + (R^T U)^\dagger \chi_2 \right)
\right]
\,,
\label{NLsM-Lag-Op2-mass}
\eeq
where the spurion fields $\chi_1$ and $\chi_2$ transform under the $G$-symmetry as 
\beq
\chi_1 \to g_{\tilde{3L}} \cdot \chi_1 \cdot g^\dagger_{\tilde{3L}} 
\,,\quad 
\chi_2 \to g_{\tilde{3L}} \cdot \chi_2 \cdot g^\dagger_{1R} 
\,.  
\eeq
The $G$-symmetry is explicitly broken 
when the spurion fields acquire the vacuum expectation values, 
\beq
\vev{\chi_1} = \vev{\chi_2} = \Sigma 
\,, 
\eeq
so that the $c_1$ and $c_2$ terms break the $G$ down to 
$U(2)_{q^3_L} \times U(1)_{\chi_L} \times U(1)_{\chi_R}$ 
and $U(2)_{q^3_L} \times U(1)_{V=\chi_R+\chi_L}$. 
The coefficients $c_1$ and $c_2$ in Eq.(\ref{NLsM-Lag-Op2-mass}) 
are fixed so as to give the mass formula between two Top-Mode Pseudos 
at the tree-level of the perturbation with respect to 
the explicit breaking couplings~\cite{Fukano:2013aea}, 
\beq 
m^{(0)}_{h^0_t} 
=
m^{(0)}_{A^0_t} \sin\theta
\,, 
\label{mass-ht0}
\eeq 
so that 
\beq
c_1 = - \frac{1}{2} \left( m^{(0)}_{A^0_t} \right)^2 
\,, \qquad 
c_2 = \frac{1}{2} \left( m^{(0)}_{A^0_t} \right)^2 \cos \theta 
\,.\nonumber
\eeq 

To describe interactions between fermions and the Top-Mode Pseudos, 
we may add the top- and $\chi$-quark sectors 
to the nonlinear Lagrangian~\cite{Fukano:2013aea},
\beq
{\cal L}^{t,t'}_{\rm yuk.}
=
-
\frac{f}{\sqrt{2}}
\left[ 
y \bar{\psi}_L (R^T U) \psi_R 
+
y_{\chi t}\bar{\psi}_L (\chi_1 R^T U \chi_3) \psi_R
+ \text{h.c.}
\right]
\,,  
\label{NLsM-Lag-Op2-withtopmass}
\eeq
where $\psi_R=(q^3_R, \chi_R)^T=((t_R, b_R), \chi_R)^T$.  
The spurion fields $\chi_1$ and $\chi_3$ have been introduced 
in Eq.(\ref{NLsM-Lag-Op2-withtopmass}) and transform as 
\beq
\chi_1 \to g_{3 L} \cdot \chi_1 \cdot g^\dagger_{3L} 
\,, \quad
\chi_3\to g_{1R} \cdot \chi_3 \cdot g^\dagger_{1R}
\,,  
\eeq
so that the Lagrangian Eq.(\ref{NLsM-Lag-Op2-withtopmass}) is invariant 
under the $G$ symmetry, 
$U(3)_{\tilde{\psi}_L} \times U(1)_{\chi_R}$, and $U(2)_{q^3_R}$ symmetry. 
These symmetries are explicitly broken by the vacuum expectation values of the spurion fields, 
\beq 
\vev{\chi_1} = \Sigma 
\,,\quad 
\vev{\chi_3} = \lambda_4 
\,, 
\eeq
in which the $\vev{\chi_1}$ breaks the $U(3)_{\psi_L}$ symmetry down to 
$U(2)_{\psi_L} \times U(1)_{\chi_L}$ 
and the $\vev{\chi_3}$ breaks the $U(2)_{q_R^3}\times U(1)_{\chi_R}$ down to 
$U(1)_{\chi_R=t_R}$. 
Equation (\ref{NLsM-Lag-Op2-withtopmass}) gives the fermion mass matrix of seesaw type 
to be diagonalized by an orthogonal rotation as 
\beq
-\bpm \bar{t}_L & \bar{\chi}_L \epm_g
\bpm 0 & m_{t\chi} \\ \mu_{\chi t} & m_{\chi \chi} \epm
\bpm t_R \\ \chi_R \epm_g + \text{h.c.}
= 
-\bpm \bar{t}_L & \bar{t}'_L \epm_m
\bpm m_t & 0 \\0 & m'_t \epm
\bpm t_R \\ t'_R \epm _m+ \text{h.c.}
\,,\label{tt'-mass-matrix}
\eeq 
where $m_{t \chi}, \mu_{\chi t}$ and $m_{\chi\chi}$ can be expressed 
as a function of $y, y_{\chi t}$ and $f$, 
and the subscripts $g$ and $m$ imply the gauge (current) and mass eigenstates, respectively, 
which are related by the orthogonal rotation, 
\beq
\bpm t_{L} \\[1ex] t'_{L} \epm_m
=
\bpm c^t_{L} & -s^t_{L} \\[1ex] s^t_{L} & c^t_{L}\epm
\bpm t_{L} \\[1ex] \chi_{L} \epm_g
\quad , \quad
\bpm t_{R} \\[1ex] t'_{R} \epm_m
=
\bpm -c^t_{R} & s^t_{R} \\[1ex] s^t_{R} & c^t_{R}\epm
\bpm t_{R} \\[1ex] \chi_{R} \epm_g
\,,\label{rotate-ttprime}
\eeq
with $c^t_{L(R)} \equiv \cos \theta^t_{L(R)}$ and $s^t_{L(R)} \equiv \sin \theta^t_{L(R)}$.   
The explicit expressions for the mass eigenvalues $(m_t, m_{t'})$ can be found 
in Ref.\cite{Fukano:2013aea} 
and we will not display them here 
since they are irrelevant to the present study. 
We shall take $y_{\chi t}/y < 1$ in order to realize 
the top-seesaw mechanism~\cite{Dobrescu:1997nm,Chivukula:1998wd}, 
which turns out to be consistent also with the constraint on the $t'$-quark mass  
from the electroweak precision tests~\cite{Fukano:2013aea}. 
The angles $c^t_{L(R)}$  and $s^t_{L(R)}$ can then be expanded in powers of $y_{\chi t}/y$ 
to be expressed up to ${\cal O}(y^3_{\chi t}/y^3)$ as
\beq 
c^t_L 
&=& 
\frac{1}{\sqrt{2}} \left[ 
1 + \frac{m^2_{\chi \chi} -m^2_{t\chi} + \mu^2_{\chi t}}{m^2_{t'} - m^2_t}
\right]^{1/2}
\simeq
\cos \theta \left[
1+ \frac{y^2_{\chi t}}{y^2} \cos^2\theta \sin^2\theta
\right]
\,,\label{def-cLt}
\\
s^t_L 
&=& 
\frac{1}{\sqrt{2}} \left[ 
1 - \frac{m^2_{\chi \chi} -m^2_{t\chi} + \mu^2_{\chi t}}{m^2_{t'} - m^2_t}
\right]^{1/2}
\simeq
\sin \theta \left[ 
1- \frac{y^2_{\chi t}}{y^2} \cos^4\theta 
\right]
\,,\label{def-sLt}
\\
c^t_R
&=& 
\frac{1}{\sqrt{2}} \left[ 
1 + \frac{m^2_{\chi \chi} + m^2_{t \chi}  -\mu^2_{\chi t}}{m^2_{t'} - m^2_t}
\right]^{1/2}
\simeq
1 - \frac{1}{2} \frac{y^2_{\chi t}}{y^2} \cos^4\theta
\,,\label{def-cRt}
\\
s^t_R
&=& 
\frac{1}{\sqrt{2}} \left[ 
1 - \frac{m^2_{\chi \chi} + m^2_{t \chi}  -\mu^2_{\chi t}}{m^2_{t'} - m^2_t}
\right]^{1/2}
\simeq
\frac{y_{\chi t}}{y}\cos^2\theta 
\left[
1- \frac{1}{2}\frac{y^2_{\chi t}}{y^2} \cos^2\theta (\cos^2\theta - 2 \sin^2\theta)
\right]
\,.\label{def-sRt}
\eeq

As discussed in Ref.\cite{Fukano:2013aea}, 
the SM fermions other than the top quark are also allowed to acquire the masses 
and couple to the Top-Mode Pseudos by introducing some four-fermion interactions 
communicating with top and $\chi$ quarks, 
through the nonzero condensate $\vev{\bar{\chi}_R t_L}$, 
without invoking other condensations like bottom condensation. 
In terms of the nonlinear sigma model, 
such four-fermion terms can be replaced with the Yukawa interaction terms
\beq 
{\cal L}^{\rm others} _{\rm yuk.}
&=& - \frac{f}{\sqrt{2}}  
\left[ 
 \sum_{\alpha=1,2}
y_{u^{\alpha}} \bar{\psi}_L^{\alpha} 
\left(  
\chi_4 R^T U \chi_5    
\right) 
\psi_R^\alpha 
- 
 \sum_{\alpha=1,2,3}
\, i y_{d^\alpha} \bar{\psi}_L^{\alpha} 
\left(  
\chi_6 R^T U \chi_7   
\right)
\psi_R^\alpha
\right.
\nonumber \\ 
&&
\left.
\hspace*{7ex}
+ 
 \sum_{\alpha=1,2,3}
\, i y_{l^\alpha} \bar{l}_L^{\alpha} 
\left(  
\chi_6 R^T U \chi_7   
\right)
\l_R^\alpha
+ {\rm h.c.} 
\right] 
\,, \label{lag:other-yukawa}
\end{eqnarray}
where $\psi^\alpha_{L,R} = (q^\alpha_{L,R}, 0)^T = ((u^\alpha_{L,R}, d^\alpha_{L,R}), 0)^T$ 
and $l^\alpha_{L,R} = (\nu^\alpha_{L,R},e^\alpha_{L,R}, 0)^T$. 
The spurion fields $\chi_{4,5,6,7}$ have been introduced 
so as to make the Lagrangian invariant under the $G$. 
They have the vacuum expectation values: 
\beq
\vev{\chi_4} 
=
\bpm
1 & 0 & 0 \\ 
0 & 1 & 0 \\ 
0 & 0 & 0  
\epm
\,,\qquad 
\vev{\chi_5} = \lambda_4 
\,,\qquad
\vev{\chi_6} 
= 
\lambda_2 
\,,\qquad 
\vev{\chi_7} 
= 
\lambda_6 
\,. 
\eeq
From Eq.(\ref{lag:other-yukawa}), 
the light SM fermions get masses as 
$m_{u^\alpha} = y_{u^\alpha}/\sqrt{2} v_{_{\rm EW}}, 
m_{d^\alpha} = y_{d^\alpha}/\sqrt{2} v_{_{\rm EW}}$, and 
$m_{e^\alpha} = y_{l^\alpha}/\sqrt{2} v_{_{\rm EW}}$, 
where use has been made of Eq.(\ref{def-vEW}). 
In addition, we find the Yukawa couplings to the Top-Mode Pseudos: 
\beq
{\cal L}^{\rm others}_{\rm yuk.}
\ni 
- \cos \theta \left[ 
\sum_{\alpha=1,2} \frac{m_{u^\alpha}}{v_{_{\rm EW}}} h^0_t\bar{u}^\alpha u^\alpha
+
\sum_{\alpha=1,2,3}\frac{m_{d^\alpha}}{v_{_{\rm EW}}} h^0_t\bar{d}^\alpha d^\alpha
+
\sum_{\alpha=1,2,3}\frac{m_{e^\alpha}}{v_{_{\rm EW}}} h^0_t\bar{e}^\alpha e^\alpha
\right]
\,. \label{yukawa-no-top}
\eeq 
Note the absence of Yukawa couplings to 
the CP-odd Top-Mode pseudo $A^0_t$ for the light SM fermions, 
due to the orthogonality of the associated $A^0_t$ current to the SM light fermion currents. 
This implies that the $A^0_t$ cannot be produced through the Drell-Yan process 
or decay to $\tau^+\tau^-$ and $b\bar{b}$. 
The former in particular leads to no interference 
in searches for the SM-like Higgs produced with the $Z$ boson, 
i.e. $q\bar{q} \to Zh$, 
in sharp contrast to other CP-odd Higgs bosons such as those in the 2HDM. 
 
Thus, the phenomenological Lagrangian for the Top-Mode Pseudos is given by 
\beq
{\cal L}
=
{\cal L}_{\text{NL$\sigma$M}}
+
\Delta {\cal L}_{\text{NL$\sigma$M}}
+
{\cal L}^{t,t'}_{\rm yuk.}
+
{\cal L}^{\rm others}_{\rm yuk.} \, \text{ [Eq.(\ref{yukawa-no-top})]}
\label{TMP-EFT}
\eeq
In the following sections, 
we will employ the LHC phenomenologies of the Top-Mode Pseudos 
based on the Lagrangian Eq.(\ref{TMP-EFT}). 

Before proceeding to the LHC phenomenology, 
we shall remark on the radiative corrections to the Top-Mode pseudo masses, 
arising as the next-to-leading-order terms in the perturbation with respect to 
the explicit breaking parameters 
$c_{1,2}, y_{\chi t}$ in 
$\Delta {\cal L}_{\textrm{NL}\sigma{\rm M}}$ and ${\cal L}^{t,t'}_{\rm yuk.}$. 
Of these corrections at the one-loop order,  
the top and $t'$ loop arising as terms of 
${\cal O}(y^2_{\chi t}) \sim {\cal O}(m^2_t/v^2_{_{\rm EW}})$ 
will give the most sizable contributions~\cite{Fukano:2013aea}. 
However, we may integrate out the $t'$ quark so as not to incorporate 
the loop contribution to the Top-Mode Pseudo masses;
recall the relationship between the masses of 
the $\chi$-quark and the heavy Higgs, which is set by the usual 
formula derived from the NJL dynamics, $m_{H^0_t} = 2 m_{\tilde{\chi} \chi}$, 
where $m_{\tilde{\chi} \chi} = \sqrt{m^2_{\chi\chi} + m^2_{t \chi}}$. 
As it will turn out, 
the $t'$-quark is required to be almost composed from 
the $\chi$ quark with $s^t_L$ in Eq.(\ref{def-sLt}) 
being approximated to be 
$\sim \sin\theta < 0.3$ to be consistent with 
the Higgs coupling measurement at the LHC Run-I 
[see Eqs.(\ref{constraint-cos}) and (\ref{constraint-cos:2})]. 
Hence, we may take $m_{t'} \sim m_{\tilde{\chi}\chi}$, 
which is quite close to the heavy Higgs mass scale, i.e.,
 the cutoff scale of the nonlinear sigma model. 
We may thus integrate out the $t'$ quark 
and take the $t'$ quark mass to be the cutoff 
of the effective theory. 
In that case, the mass shifts of two Top-Mode Pseudo masses are given by~\cite{Fukano:2013aea}
\beq
m^2_{h^0_t} 
&=&
\left( m^{(0)}_{h^0_t} \right)^2
- 
\frac{3 m^2_{t'}}{8\pi^2} \left( \frac{\sqrt{2} m_t}{v_{_{\rm EW}}} \right)^2 
\frac{1 - 6 \cos^2\theta + 6 \cos^4 \theta}{\cos^2\theta} 
\left[ 
1 +{\cal O}\left( \frac{y^2_{\chi t}}{y^2} \right)
\right]
\,, \label{t-prime-integ-mh2}
\\
m^2_{A^0_t} 
&=&
\left( m^{(0)}_{A^0_t} \right)^2
- 
\frac{3 m^2_{t'}}{8\pi^2} \left( \frac{\sqrt{2} m_t}{v_{_{\rm EW}}} \right)^2 
\frac{(1-\cos^2\theta)(2-7\cos^2\theta +6\cos^4\theta)}{2\cos^2\theta}
\left[ 
1 +{\cal O}\left( \frac{y^2_{\chi t}}{y^2} \right)
\right]
\,.\label{t-prime-integ-mA2}
\eeq
From these, 
we see that 
setting the tHiggs mass at one-loop level  
$m_{h^0_t} =126\,\GeV$ requires the tree-level mass to be 
\beq 
m^{(0)}_{h^0_t} \simeq 230\,\GeV
\,, \label{hmass:tree}
\eeq 
for $m_{t'} \simeq 1.2\,\TeV$, $y_{\chi t}/y \simeq 0.7$, 
and $\cos \theta \sim 0.97$, which is consistent with 
the Higgs coupling measurement, as will be seen later, 
and the $S,T$ parameter constraint~\cite{Fukano:2013aea}.  
Note, on the other hand, that  
the $A^0_t$ mass is almost stable against the top quark loop for $\cos\theta \sim 1$: 
\beq 
m_{A^0_t} \simeq m^{(0)}_{A^0_t}
\,. \label{Amass:tree}
\eeq

\section{CP-even Top-Mode Pseudo ($\text{t}$Higgs $h^0_t$) at The LHC}
\label{Ph-ht0}

In this section, 
we discuss the coupling properties of the tHiggs $h^0_t$ with $m_{h^0_t} = 126\,\GeV$ 
in comparison with the currently available Higgs search data at the LHC.
We further place the limit on the mass of $A^0_t$ by using the mass relation 
between two Top-Mode Pseudos Eq.(\ref{mass-ht0}). 

\subsection{tHiggs coupling properties} 

After the $t'$ quark is integrated out 
by assuming $m_{t'}\gg m_{t,h^0_t,A^0_t} $, 
the relevant tHiggs interaction terms are read off from the Lagrangian Eq.(\ref{TMP-EFT}),
\beq
{\cal L}_{h^0_t}
\!\!\!&=&\!\!\!
g_{hVV} \frac{v_{_{\rm EW}}}{2}  
\left( g^2 h^0_t W^+_\mu W^{-\mu} 
+ 
\frac{g^2 + g'^2}{2} h^0_t Z_\mu Z^\mu \right)
\nonumber
\\[1ex]
&&
-g_{htt} \frac{m_t}{v_{_{\rm EW}}} h^0_t \bar{t}t 
-g_{hbb} \frac{m_b}{v_{_{\rm EW}}} h^0_t \bar{b}b 
-g_{h\tau\tau} \frac{m_\tau}{v_{_{\rm EW}}} h^0_t\bar{\tau}\tau
\,,\label{tHiggs-Lag}
\eeq
where
\beq
&&
g_{hVV} 
= g_{hbb} 
= g_{h\tau\tau}
= \cos \theta
\,,\label{gTM-hVVhhff}\\
&&
g_{htt} 
=
\frac{v_{_{\rm EW}}}{m_t}\frac{y}{\sqrt{2}} 
\left[ 
(c^t_L \cos \theta  + s^t_L \sin \theta ) s^t_R 
- 
s^t_Lc^t_R \sin\theta \left( \frac{y_{\chi t}}{y} \right)
\right]
= 
\frac{2\cos^2\theta -1}{\cos\theta} + {\cal O} \left( \frac{y^2_{\chi t}\sin^2\theta}{y^2} \right)
\,.\label{gTM-htt}
\eeq
We may further incorporate the tHiggs couplings to $gg$ and $\gamma\gamma$ 
generated at the one-loop level,
\beq
{\cal L}^{gg,\gamma\gamma}_{h^0_t}
=
\left( g_{h gg}  + \Delta g^{(t')}_{h gg} \right)
\frac{\alpha_s}{16\pi v_{_{\rm EW}}} h^0_t G^{\mu \nu}G_{\mu \nu} 
+
\left( g_{h\gamma\gamma} + \Delta g^{(t')}_{h \gamma\gamma} \right) 
\frac{\alpha}{8\pi v_{_{\rm EW}}} h^0_t F^{\mu \nu}F_{\mu \nu}
\,,\label{hgg-hgammgamma}
\eeq
where $\alpha_s \equiv g^2_s/(4\pi)$ with $g_s$ being the $SU(3)_c$ gauge coupling 
and $\alpha \equiv e^2/(4\pi)$ with $e$ being the electromagnetic coupling. 
The coefficients $g_{hgg}, \Delta g^{(t')}_{hgg}$, $g_{h\gamma\gamma}$, 
and $\Delta g^{(t')}_{h\gamma\gamma}$ in Eq.(\ref{hgg-hgammgamma}) 
are 
\beq
g_{hgg}
&=&
\sum_f g_{hff} A^h_{1/2}(\tau_f)
\,,\label{ghgg}
\\
\Delta g^{(t')}_{hgg}
&=& 
\frac{4}{3} \sin^2\theta \left(  \frac{y_{\chi t}}{y} \right)^2 \left[ 1 + {\cal O}\left( \frac{y_{\chi t}}{y} \right)^2 \right]
\,, \label{t-prime-hgg} \\ 
g_{h\gamma\gamma}
&=&
g_{hVV} A_1(\tau_W)
+
\sum_f N^{(f)}_c Q^2_f g_{hff} A^h_{1/2}(\tau_f)
\,,\label{ghgammagamma}
\\ 
\Delta g^{(t')}_{h \gamma\gamma}
&=& 
\frac{16}{9} \sin^2\theta \left(  \frac{y_{\chi t}}{y} \right)^2 \left[ 1 + {\cal O}\left( \frac{y_{\chi t}}{y} \right)^2 \right] 
\,, \label{t-prime-hgammagamma}
\eeq
where $N^{(f)}_c = 3(1)$ for quarks (leptons), $\tau_i \equiv 4m^2_i/m^2_h$, 
and the functions $A_1(x)$ and $A^h_{1/2}(x)$ are defined as
\beq
A_1(x)&=&2+3x+3x(2-x)f(x)
\,,\\
A^h_{1/2}(x)&=&
2x[1+(1-x)f(x)] 
\,,\\
f(x) &=&
\begin{cases}
\left[ \arcsin (1/\sqrt{x})\right]^2 & \text{for $x>1$}
\\[1ex]
-\frac{1}{4} \left[ \ln \dfrac{1+\sqrt{1-x}}{1-\sqrt{1-x}} - i\pi\right]^2 & \text{for $x\leq 1$}
\end{cases}
\,.\label{def-higgsdecay-fn}
\eeq
Note the terms in Eqs.(\ref{t-prime-hgg}) 
and (\ref{t-prime-hgammagamma}) corresponding to the nondecoupling contributions 
from integrating out the $t'$ quark. 
However, 
the $t'$ contributions are numerically negligible since 
the overall factor $\sin^2\theta$ turns out to be constrained by 
the Higgs coupling measurement as $\sin\theta \lesssim 0.2 - 0.4$ 
[see Eqs.(\ref{constraint-cos}) and (\ref{constraint-cos:2}) 
and this is also the case for $A^0_t$ as will be seen later].
Thus, the $h^0_t$-$g$-$g$ and $h^0_t$-$\gamma$-$\gamma$ couplings 
approximately become the same as those of the SM Higgs boson. 
Note that the CP-odd Top-Mode Pseudo $A^0_t$ is necessarily heavier than 
the tHiggs [see Eqs.(\ref{mass-ht0}), (\ref{t-prime-integ-mh2}) and (\ref{t-prime-integ-mA2})]; 
hence, the tHiggs cannot decay to $A^0_t$. 
Therefore, 
both the production cross sections and decay properties of $h^0_t$ are 
almost the same as those of the SM Higgs boson, 
up to some size of a deviation controlled by 
a coupling parameter $\cos\theta$.

\subsection{Fitting the tHiggs couplings to the LHC Run-I data}  

\begin{figure}[htb]
\begin{center}
\begin{tabular}{cc}
{
\begin{minipage}[t]{0.4\textwidth}
\includegraphics[scale=0.6]{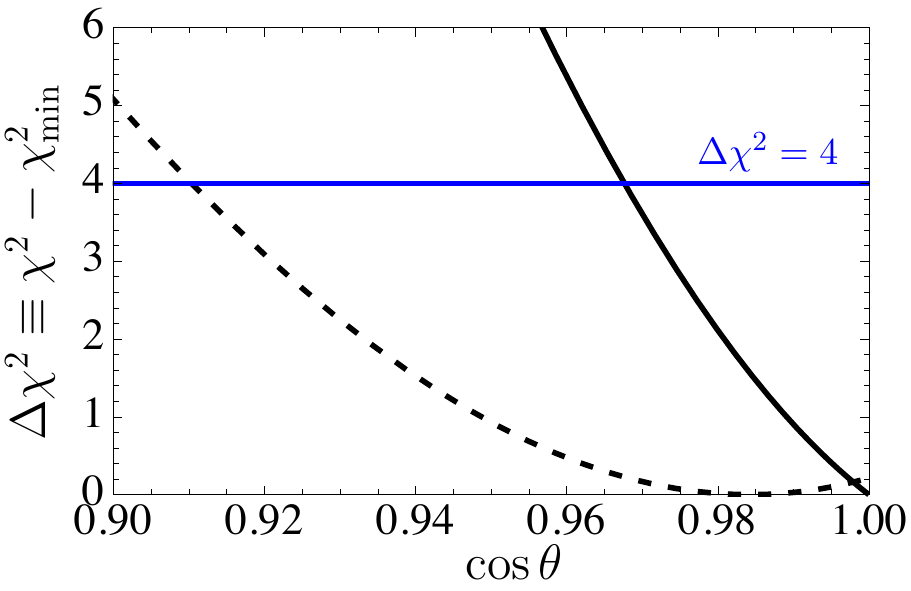} 
\end{minipage}
}
&
{
\begin{minipage}[t]{0.4\textwidth}
\includegraphics[scale=0.65]{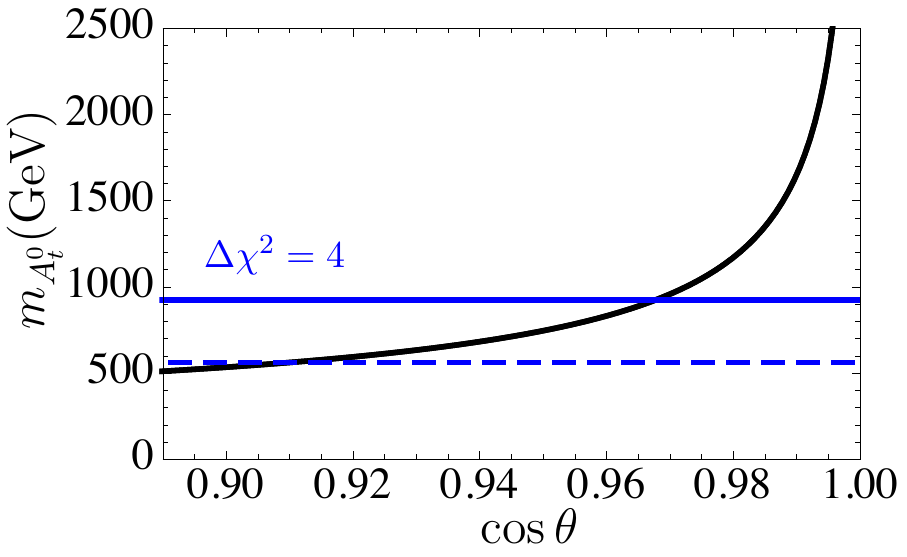} 
\end{minipage}
}
\end{tabular}
\caption[]{
The plots of $\Delta \chi^2 \equiv \chi^2 - \chi^2_{\rm min}$ (left panel) 
and $m_{A^0_t}$ (right panel) as a function of $\cos\theta$ 
together with the $95\%\cl$ upper limits represented 
by the $\Delta \chi^2=4$ lines. 
In the left panel, 
the dashed and solid curves have, respectively, been created 
by the $\chi^2$ goodness-of-fit tests with and without 
the incorporation of the correlation mentioned in the text. 
In the right panel, 
the cross points made of 
the black-solid curve with blue-dashed and blue-solid horizontal lines 
correspond to the $95\%\cl$ lower limits given 
in Eq.(\ref{constraint-cos}) [or Eq.(\ref{constraint-mAt})] 
and 
Eq.(\ref{constraint-cos:2}) [or Eq.(\ref{constraint-mAt:2})], 
respectively.
\label{TMP-constraint}} 
\end{center}
\end{figure}

The ATLAS~\cite{ATLAS-CONF-2013-012,Aad:2013wqa,
ATLAS-CONF-2013-030,ATLAS2013108,TheATLAScollaboration:2013lia}  
and 
CMS~\cite{CMS-PAS-HIG-13-001,CMS-PAS-HIG-13-002,Chatrchyan:2013iaa,CMS:2013yea,
CMS-PAS-HIG-13-004,Chatrchyan:2013zna}
collaborations have provided the signal strengths $\hat{\mu}$ of the 126 GeV Higgs boson 
for each decay channel, which are classified by the production processes; 
gluon-gluon fusion (ggF) plus top quark associate productions (t$\bar{\text{t}}$H),
$\hat{\mu}(\text{ggF+t$\bar{\text{t}}$H})$;
and 
vector boson fusion (VBF) plus vector boson associate productions (VH), 
$\hat{\mu}(\text{VBF+VH})$~\footnote{
As noted around Eq.(\ref{yukawa-no-top}), 
the CP-odd Top-Mode Pseudo $A^0_t$ does not interfere the Higgs search in the 
VH channel because of no couplings to light quarks, in sharp 
contrast to other CP-odd Higgs like those in the 2HDM. 
This makes it possible to directly quote the Higgs coupling data from the VH channel 
to constrain solely the tHiggs without resonance contributions of CP-odd Higgs bosons. 
}. 
\begin{table}[t]
\begin{center}
\begin{tabular}{| c | c  | c | c | c || c |}
\hline
Decay channel 
&  $\hat{\mu}(\text{ggF+t$\bar{\text{t}}$H})$ 
&  $\hat{\mu}(\text{VBF+VH})$ 
& $\Delta\mu(\text{ggF+t$\bar{\text{t}}$H})$ 
& $\Delta\mu(\text{VBF+VH})$ 
& Ref.
\\
\hline
\hline
\parbox[c][3ex][c]{0ex}{}
$\gamma\gamma$ (ATLAS)
& $1.6$ & $1.7$
& $0.25$ & $0.63$ 
& \cite{ATLAS-CONF-2013-012}
\\
\hline 
\parbox[c][3ex][c]{0ex}{}
$ZZ^*$ (ATLAS) 
& $1.8$ & $1.2$ 
& $0.35$ & $1.30$
& \cite{Aad:2013wqa}
\\
\hline
\parbox[c][3ex][c]{0ex}{}
$WW^*$ (ATLAS) 
& $0.82$  & $1.66$
& $0.36$ & $0.79$
& \cite{ATLAS-CONF-2013-030}
\\ 
\hline
\parbox[c][3ex][c]{0ex}{}
$\tau\tau$ (ATLAS)
& $1.1$  & $1.6$ 
& $1.16$ & $0.75$
& \cite{ATLAS2013108}
\\ 
\hline
\parbox[c][3ex][c]{0ex}{}
$b\bar{b}$ (ATLAS)
& --  &  $0.2$
& -- & $0.64$
& \cite{TheATLAScollaboration:2013lia}
\\ 
\hline
\hline
\parbox[c][3ex][c]{0ex}{}
$\gamma\gamma$ (CMS)
& $0.52$ & $1.48$
& $0.60$ & $1.33$
& \cite{CMS-PAS-HIG-13-001}
\\
\hline 
\parbox[c][3ex][c]{0ex}{}
$ZZ^*$ (CMS)
& $0.9$ & $1.0$
& $0.45$ & $2.35$
& \cite{CMS-PAS-HIG-13-002}
\\
\hline
\parbox[c][3ex][c]{0ex}{}
$WW^*$ (CMS)
& $0.72$ & $0.62$
& $0.37$  & $0.53$
& \cite{Chatrchyan:2013iaa,CMS:2013yea}
\\ 
\hline
\parbox[c][3ex][c]{0ex}{}
$\tau\tau$ (CMS)
& $1.07$ & $0.94$
& $0.46$  & $0.41$ 
& \cite{CMS-PAS-HIG-13-004}
\\ 
\hline
\parbox[c][3ex][c]{0ex}{}
$b\bar{b}$ (CMS)
& --  &  $1.0$
& -- & $0.5$
& \cite{Chatrchyan:2013zna}
\\ 
\hline
\end{tabular}
\caption{
The best-fit signal strengths 
$\hat{\mu}(\text{ggF+t$\bar{\text{t}}$H})$ and $\hat{\mu}(\text{VBF+VH})$ 
reported from the Higgs search  
at the ATLAS~\cite{ATLAS-CONF-2013-012,Aad:2013wqa,
ATLAS-CONF-2013-030,ATLAS2013108,TheATLAScollaboration:2013lia}  
and 
CMS~\cite{CMS-PAS-HIG-13-001,CMS-PAS-HIG-13-002,Chatrchyan:2013iaa,CMS:2013yea,
CMS-PAS-HIG-13-004,Chatrchyan:2013zna} 
experiments.
As for the $WW^*$ channel (CMS), 
the value of $\hat{\mu}(\text{ggF+t$\bar{\text{t}}$H})$ is taken from Ref.\cite{Chatrchyan:2013iaa}, 
and 
the value of $\hat{\mu}(\text{VBF+VH})$ is from Ref.\cite{CMS:2013yea}. 
The value of $\hat{\mu}(\text{ggF+t$\bar{\text{t}}$H})$ for the $\tau\tau$ channel (CMS) 
is quoted from the one-jet result in Ref.\cite{CMS-PAS-HIG-13-004}.
} 
\label{mu-LHC}
\end{center}
\end{table}%
In Table.\ref{mu-LHC}, 
we present the signal strengths reported by the ATLAS and CMS collaborations. 
By using them, 
we construct a simple $\chi^2$ function as 
\beq
\chi^2(\theta)
\equiv
\sum_X
\sum_{C,i,j}
\left(
\frac{\mu^X_i(\theta)-\hat{\mu}^X_{C,i}}{\Delta \mu^X_{C,i}}
\right)^2
\,,\label{chi2-def-tHiggs}
\eeq
where 
$\mu^X_i(\theta)$ implies the signal strength of tHiggs 
for each production channel $i,j \in \{ \text{ggF+t$\bar{\text{t}}$H,VBF+VH}\}$ 
and each decay channel $X \in \{ \gamma\gamma,ZZ^*,WW^*,\tau\tau,b\bar{b}\}$. 
We use $G_F = 1.166 \times 10^{-5}\,\GeV^{-2}$, 
$m_Z = 91.188\,\GeV$, $\alpha = 1/137$, $m_t = 173.1\,\GeV$, 
and $\alpha_s(m_Z) = 0.118$ as inputs \cite{Beringer:1900zz}. 
Then, the signal strength of the tHiggs depends only on $\cos\theta$, 
which parametrizes couplings between tHiggs and SM particles 
as seen from Eqs.(\ref{tHiggs-Lag}), (\ref{gTM-hVVhhff}), and (\ref{gTM-htt}).
$\hat{\mu}^X_{C,i}$ is the value of the best-fit signal strength of the 126 GeV Higgs 
for each production $(i)$ and decay channel $(X)$ reported by 
the experiments $C \in\{ \text{ATLAS,CMS}\}$. 
We may take into account the next-to-leading-order corrections to 
the ggF process arising from QCD, the so-called K-factor, 
for the CP-even scalar~\cite{Djouadi:2005gi}, $K^g_h = 1 + (215/12) \alpha_s(m_h)/\pi $, 
where $\alpha_s(m_h)$ is the one-loop QCD gauge coupling at the scale $\mu = m_h$. 
In the left panel of Fig.~\ref{TMP-constraint} (solid curve), 
we show $\Delta \chi^2 \equiv \chi^2 - \chi^2_{\rm min}$ as a function of $\cos \theta$, where
$\chi^2_{\rm min} = 13.5$ at $\cos\theta= 1$ for the number of degrees of freedom being 18. 
From the left panel, we find the $95\%\cl$ allowed region for $\cos\theta$:
\beq
0.97 \leq \cos \theta \leq 1
\,.\label{constraint-cos}
\eeq
Using the mass relation between $m^{(0)}_{h^0_t}$ and $m^{(0)}_{A^0_t}\simeq m_{A^0_t}$ 
given in Eq.(\ref{mass-ht0}) with Eq.(\ref{hmass:tree}) taken into account, 
from Eq.(\ref{constraint-cos}), we may place an indirect bound on the mass of $A^0_t$. 
In the right panel of Fig.~\ref{TMP-constraint}, 
we plot $m_{A^0_t}$ as a function of $\cos\theta$,  
from which the $95\%\cl$ allowed region of $m_{A^0_t}$ (horizontal solid line) reads  
\beq
m_{A^0_t} \geq 923 \,\GeV
\,.\label{constraint-mAt}
\eeq

We may take into account the correlation between 
$\hat{\mu}(\text{ggF+t$\bar{\text{t}}$H})$ and $\hat{\mu}(\text{VBF+VH})$, 
which can be read off from Ref.\cite{Belyaev:2013ida}, 
though it is not in public.
The $95\%\cl$ allowed region in Eqs.(\ref{constraint-cos}) and (\ref{constraint-mAt}) 
would change to
\beq
0.91 \leq \cos \theta \leq 1 
\,, \label{constraint-cos:2}
\eeq
for $\cos \theta$ and (see the horizontal dashed line in the right panel of Fig.~\ref{TMP-constraint})
\beq
m_{A^0_t} \geq 563 \,\GeV
\,, \label{constraint-mAt:2}
\eeq
for $m_{A^0_t}$. 
Thus, the incorporation of the correlation would make 
the lower bounds on $(\cos\theta, m_{A^0_t})$ milder. 
In the present study, therefore, 
we shall explore the LHC phenomenology of $A^0_t$ by scanning 
the mass in the high-mass range [Eq.(\ref{constraint-mAt})] 
and the low-mass range [Eq.(\ref{constraint-mAt:2})]. 

\section{CP-odd Top-Mode Pseudo ($A^0_t$) at the LHC}
\label{Ph-At0}

In this section, 
we explore the LHC phenomenologies of $A^0_t$ 
to compare the predicted $A^0_t$ signals in several decay channels with  
the currently available data searching for new resonances provided from LHC Run-I.

\subsection{$A^0_t$ coupling properties}  

We start with reading off the relevant $A^0_t$ interaction terms from 
the Lagrangian Eq.(\ref{TMP-EFT}):
\beq
{\cal L}_{A^0_t}
\!\!\!&=&\!\!\!
-i \left( \frac{\sin^3\theta}{\cos\theta}\right)
\frac{m_t }{v_{_{\rm EW}}} 
A^0_t \bar{t}\gamma_5t
\nonumber\\
&&
-\frac{3 \sin\theta \cos^2\theta}{4v_{_{\rm EW}}} 
\left[ 
z^0_t \partial_\mu A^0_t \partial^\mu h^0_t 
- 
h^0_t \partial_\mu A^0_t \partial^\mu z^0_t 
- 
2 \left( \left(m^{(0)}_{A^0_t} \right)^2 \sin^2\theta \right) A^0_t h^0_t z^0_t 
\right] 
\nonumber\\
&&
+
\frac{3\sin^3\theta}{4 v_{_{\rm EW}}} 
\left[ 
A^0_t \partial_\mu z^0_t \partial^\mu h^0_t 
-
h^0_t \partial_\mu A^0_t \partial^\mu z^0_t 
\right] 
\,.\label{CP-odd-TMP-Lag}
\eeq
Note that 
all the $A^0_t$ couplings vanish 
when $\sin\theta=v_{_{\rm EW}}/f \simeq m^{(0)}_{h^0_t}/m_{A^0_t}$ is sent to zero 
with $v_{_{\rm EW}}\simeq 246\,\GeV$ or $m^{(0)}_{h^0_t} \simeq 230\,\GeV$ fixed. 
This should be so 
since $A^0_t$ decouples from the low-energy effective theory 
in this limit, i.e., $m_{A^0_t} \simeq m^{(0)}_{A^0_t} \to \infty$. 
The overall suppression by this $\sin\theta$ also ensures 
the weakness of the $A^0_t$ couplings, 
leading to a quite small total width and giving the crucial difference between $A^0_t$ 
and 
other CP-odd scalars as in the MSSM/2HDM, 
as will be discussed below. 

Note also the absence of couplings to $ZZ$ and $WW$ 
since $A^0_t$ is orthogonal to the would-be NGBs $z^0_t$ and $w^\pm_t$; 
actually, there exist terms coupling to the longitudinal mode of $W^\pm$ 
like $A^0_t$-$w^-_t$-$w^+_t$, 
which, however, vanishes when the amplitude is evaluated at the $A^0_t$ mass on shell. 
Although not being displayed in Eq.(\ref{CP-odd-TMP-Lag}), 
the $A^0_t$-$Z$-$h^0_t$ term is also present in the Lagrangian Eq.(\ref{TMP-EFT}),  
where the transverse component of $Z$ does not contribute in the on-shell amplitude. 
This fact is closely tied with the Goldstone boson equivalence theorem. 
It thus turns out that 
the $A^0_t$ coupling to the weak gauge bosons 
relevant to the $A^0_t$ on-shell amplitude 
is allowed only by involving both the tHiggs $h^0_t$ and the longitudinal mode $Z_L\equiv z^0_t$, 
as presented in the second line of Eq.(\ref{CP-odd-TMP-Lag}). 
Similar arguments are applicable to other CP-odd Higgs bosons such as those in the MSSM/2HDM.

\subsection{Decay properties of $A^0_t$}  
\label{At0-decay}

\begin{figure}[tb]
\begin{center}
\begin{tabular}{cc}
{
\begin{minipage}[t]{0.4\textwidth}
\includegraphics[scale=0.65]{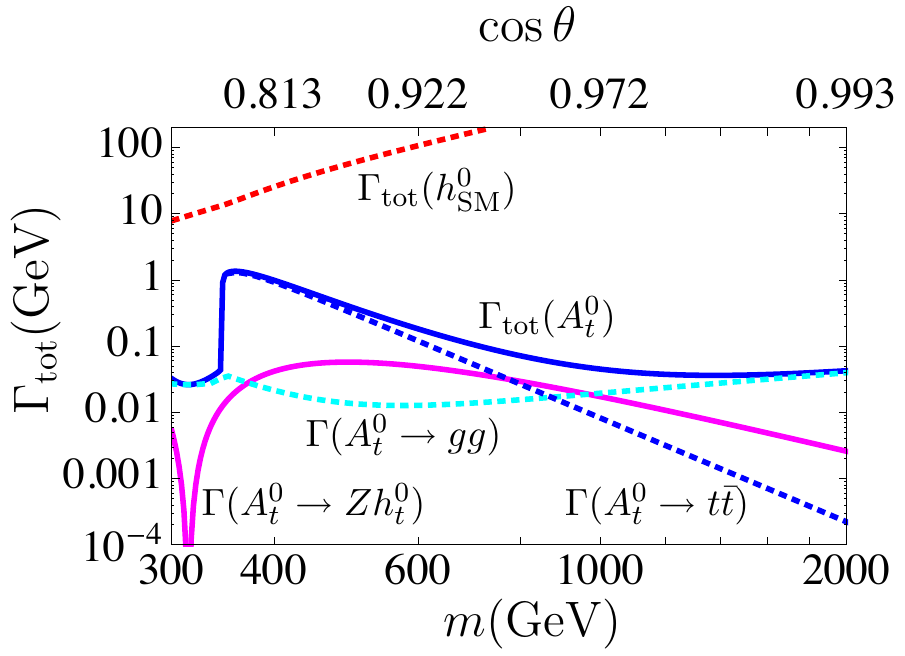} 
\end{minipage}
}
&
{
\begin{minipage}[t]{0.4\textwidth}
\includegraphics[scale=0.65]{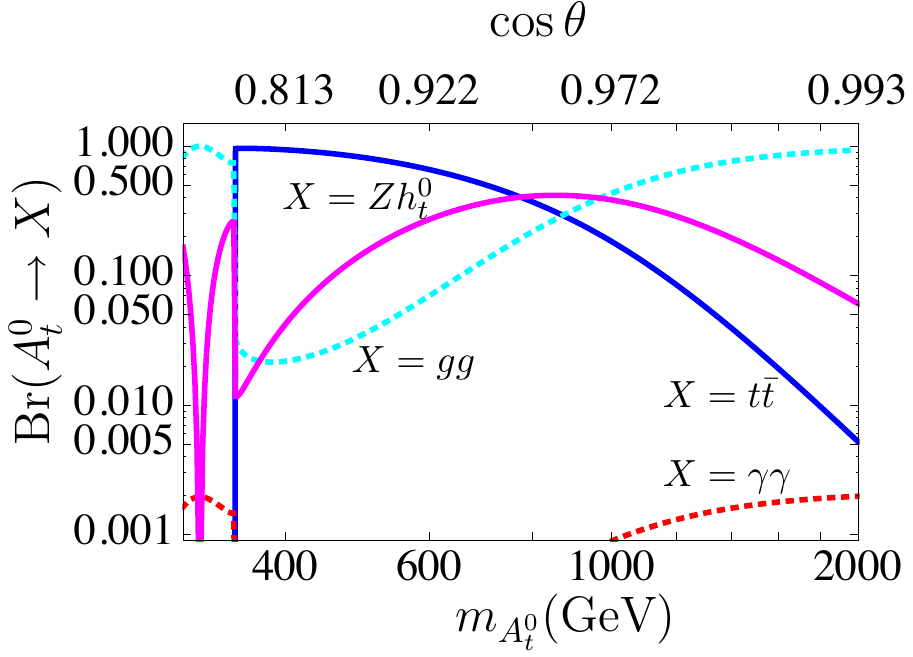} 
\end{minipage}
}
\end{tabular}
\caption[]{
Left panel: The total decay width of $A^0_t$ (blue, solid) 
and the dominant partial widths .
Right panel: The branching ratio of $A^0_t$.
Both are represented as a function of the parent particle mass 
$m=(m_{h^0_t}, m_{h_{_{\rm SM}}})$ (bottom axis) or 
$\cos \theta$ (top axis). 
In the left panel, the total decay width of the SM Higgs boson (red-dotted curve) 
is shown for comparison. 
\label{decay-At0}}
\end{center}
\end{figure}

Using Eq.(\ref{CP-odd-TMP-Lag}) and 
taking into account the loop-induced couplings to $gg$ and $\gamma\gamma$,  
we compute the partial decay widths of $A^0_t$ relevant to the two-body decay processes to obtain  
\beq 
\Gamma(A^0_t \to t\bar{t})
\!\!\!&=&\!\!\!
\frac{\sqrt{2}G_F N_c m^2_t m_{A^0_t}}{8\pi^2} 
\left(\frac{\sin^3\theta}{\cos\theta}\right)^2 \cdot \beta_A(m_t)
\,,\label{decay-At-tt}\\
\Gamma(A^0_t \to gg)
\!\!\!&=&\!\!\!
\frac{\sqrt{2}G_F \alpha^2_s m^3_{A^0_t}}{128\pi^3}
\cdot 
\left|
\left( \frac{\sin^3\theta}{\cos\theta} \right) A^A_{1/2}(\tau_t)
+
2(\sin\theta \cos\theta) 
\right|^2
\,,\label{decay-At-gg}\\
\Gamma(A^0_t \to \gamma\gamma)
\!\!\!&=&\!\!\!
\frac{\sqrt{2}G_F \alpha^2 m^3_{A^0_t}}{256\pi^3}
\cdot 
\left|
\left( \frac{\sin^3\theta}{\cos\theta} \right) N_c Q^2_t A^A_{1/2}(\tau_t)
+
\frac{8 N_c}{9} (\sin\theta \cos\theta)
\right|^2
\,,\label{decay-At-diphoton}\\
\Gamma(A^0_t \to Z_L h^0_t)
\!\!\!&=&\!\!\!
\frac{9\sqrt{2}G_F m^3_{A^0_t}}{256\pi}
\sin^2\theta \cdot 
\beta_A(m_{h^0_t})
\left[ 
\left(\sin^2\theta - \frac{m^2_{h^0_t}}{m^2_{A^0_t}} \right)
(\cos^2\theta - \sin^2\theta)
+
\frac{m^2_Z}{m^2_{A^0_t}} \cos^2\theta 
\right]^2
\,,\label{decay-At-Zh}
\eeq
where $A^A_{1/2}(x) = 2x f(x) \to 2 (x \gg 1)$ with $f(x)$ being defined 
in Eq.(\ref{def-higgsdecay-fn}) 
and 
\beq
\beta_A(m_t) \!\!\!&\equiv&\!\!\! \sqrt{1-\frac{4m^2_t}{m^2_{A^0_t}}} 
\,,\\
\beta_A(m_{h^0_t}) \!\!\!&\equiv&\!\!\!
\sqrt{
\left[ 1-\frac{\left( m_{h^0_t}-m_Z \right)^2}{m^2_{A^0_t}}\right]
\left[ 1-\frac{\left( m_{h^0_t}+m_Z \right)^2}{m^2_{A^0_t}}\right]
}
\,.
\eeq 
Note the second terms of Eqs.(\ref{decay-At-gg}) and (\ref{decay-At-diphoton}), 
coming from integrating out the $t'$ quark. 
In Fig.~\ref{decay-At0}, 
we plot the total decay width (blue-solid curve in the left panel) 
and branching ratios (right panel) of $A^0_t$ 
as a function of the parent particle mass $m=(m_{A^0_t}, m_{h_{_{\rm SM}}})$ (bottom axis) 
and $\cos \theta$ (top axis) 
where $m_{A^0_t}(\simeq m^{(0)}_{A^0_t})$ and $\cos\theta$ are related to each other 
by the mass formula Eq.(\ref{mass-ht0}) 
with $m^{(0)}_{h^0_t} \simeq 230 \,\GeV$ fixed. 
As seen from the left panel, 
$A^0_t$ is still a narrow resonance even if the mass reaches the scale over $1\,\TeV$, 
i.e., $\Gamma_{\rm tot}/m_{A^0_t} \ll 1$, 
where the $t\bar{t}$ mode rapidly damps as the mass increases. 
This happens due to the presence of the mass formula Eq.(\ref{mass-ht0}), 
which significantly affects the mass dependence of the total width for the high-mass region; 
as the mass gets larger, 
the partial decay width for the $t\bar{t}$ mode goes like $\sim 1/m^5_{A^0_t}$,  
due to the high suppression by the overall coupling 
$\sim (\sin^3 \theta)^2$ [see Eq.(\ref{decay-At-tt})], 
where $\sin \theta$ has been replaced by 
$m^{(0)}_{h^0_t}/m_{A^0_t} \simeq (230\,\GeV)/m_{A^0_t}$. 
The total width is then governed by the $gg$ mode, 
so that 
$\Gamma_{\rm tot}(A^0_t) \sim \Gamma(A^0_t \to gg) \sim m^3_{A^0_t}\tan^2\theta \sim m$,  
according to Eq.(\ref{decay-At-Zh}), 
which does not grow like $\sim m^3$ as in the case of the typical width of decays to spin-1 bosons.  
This is the salient feature of $A^0_t$ closely related to the fact that $A^0_t$ is the 
Top-Mode Pseudo partner of the Higgs $h^0_t$. 
In the left panel, 
the total width is also compared with that of the SM Higgs (red-dotted curve). 
This shows that $A^0_t$ is indeed a narrower resonance 
than the SM Higgs boson for the whole mass range. 
Note also the presence of a dip in the $A^0_t \to Zh^0_t$ channel; 
this takes place when the terms in the square bracket of Eq.(\ref{decay-At-Zh})
vanish as 
\beq
&& \left(\sin^2\theta - \frac{m^2_{h^0_t}}{m^2_{A^0_t}} \right)
(\cos^2\theta - \sin^2\theta)
+ \frac{m^2_Z}{m^2_{A^0_t}} \cos^2\theta  = 0 
\nonumber \\ 
\leftrightarrow && 
m_{A^0_t}^2 = 
\frac{\left( m^{(0)}_{h^0_t} \right)^2 
\left( m^2_Z + 2 \left( m^{(0)}_{h^0_t} \right)^2 - 2 m^2_{h^0_t} \right)
}{
\left( m^{(0)}_{h^0_t} \right)^2 - m^2_{h^0_t} + m^2_Z
}
\simeq (310\,\GeV)^2
\,, \label{dip}
\eeq
where we used the tree-level mass formula Eq.(\ref{mass-ht0}) 
with $m^{(0)}_{h^0_t} \simeq 230\,\GeV$, 
the one-loop mass $m_{h^0_t}=126\,\GeV$ and $m_Z \simeq 90\,\GeV$. 

The right panel of Fig.~\ref{decay-At0} combined with 
the indirect limits on $m_{A^0_t}$ in Eqs.(\ref{constraint-mAt}) and (\ref{constraint-mAt:2})
imply that 
the accessible decay channels of $A^0_t$ at the LHC can be 
$A^0_t \to t\bar{t},gg,Zh^0_t$:  
(i) First is the high-mass case with $m_{A^0_t} \geq 1\,\TeV$ 
indicated from the limit in Eq.(\ref{constraint-mAt})
 (hereafter, we shall call this $A^0_t$ ``high-mass $A^0_t$"),
where the $A^0_t \to gg$ mode will be 
the expected discovery channel at the LHC, 
which is accessible in a way similar to searches for new heavy bosons 
mainly decaying to gluon jets~\cite{CMS:kxa}; 
(ii) Second is the low-mass case with 
$563\,\GeV \leq m_{A^0_t} \leq 1\,\TeV$ indicated from the limit in Eq.(\ref{constraint-mAt:2}),  
where the $A^0_t \to Zh^0_t$ mode will be expected as the discovery channel at the LHC,  
which can be seen in the same way as searches for other CP-odd Higgs boson 
in the extended Higgs sector as in the MSSM/2HDM 
through the decay to $Zh^0$~\cite{CMS:2013eua}
(this $A^0_t$ will be called ``low-mass $A^0_t$"). 
Note that the $A^0_t$ still emerges as the narrow resonance in the $Zh$ channel 
(see the left panel of Fig.~\ref{decay-At0}), 
compared to other CP-odd scalars, 
like $A^0$ in the MSSM/2HDM, 
which becomes a quite broader resonance with 
the total width of ${\cal O}(100)\,\GeV$ in this mass range~\cite{Kanemura:2009mk}. 
Thus, $A^0_t$ can be distinguished from $A^0$ in the MSSM/2HDM at the LHC. 
This is mainly due to 
the presence of the intrinsic mass formula Eq.(\ref{mass-ht0}), 
which is absent in the MSSM/2HDM,  
as emphasized above. 

The predicted signals of $A^0_t$ through these decay channels 
compared with current LHC limits 
will be discussed later. 

\subsection{Production cross sections of $A^0_t$} 

\begin{figure}[htb]
\begin{center}
\begin{tabular}{c}
{
\begin{minipage}[t]{0.4\textwidth}
\includegraphics[scale=0.6]{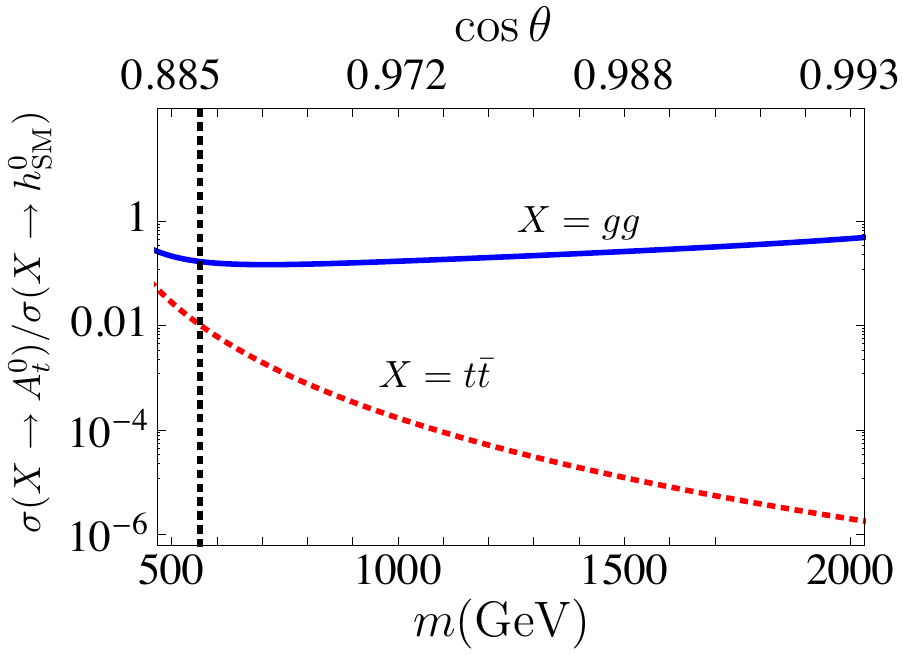} 
\end{minipage}
}
\end{tabular}
\caption[]{
The ratios of 
the production cross sections $\sigma(t\bar{t}\to A^0_t)/\sigma(t\bar{t} \to h^0_{\rm SM})$ 
(red dotted curve)
and $\sigma(gg\to A^0_t)/\sigma(gg \to h^0_{\rm SM})$ (blue solid curve) 
as a function of the produced particle mass $m=(m_{A^0_t}, m_{h_{_{\rm SM}}})$ (bottom axis) or 
$\cos \theta$ (top axis). 
The vertical dotted line corresponds to 
the indirect mass limit in Eq.(\ref{constraint-mAt:2}).
\label{cross-section}}
\end{center}
\end{figure}

The branching fraction in the right panel of Fig.~\ref{decay-At0} 
implies that at the LHC $A^0_t$ is mainly produced 
through the ggF or top quark associate process (t$\bar{\text{t}}$A) 
like the t$\bar{\text{t}}$H production for the SM Higgs. 
To make a quantitative argument, 
it is convenient to evaluate the cross section $gg/t\bar{t} \to A^0_t$ 
by normalizing it with the corresponding cross section for the SM Higgs:  
\beq 
\sigma(gg/t\bar{t} \to A^0_t)
\!\!\!&=&\!\!\!
\sigma(gg/t\bar{t} \to h^0_{\rm SM})
\times
\frac{\sigma(gg/t\bar{t}\to A^0_t)}{\sigma(gg/t\bar{t} \to h^0_{\rm SM})}
\,. \label{GFttA-A}
\eeq 
In Fig.~\ref{cross-section}, 
we plot the ratio $\sigma(gg/tt \to A^0_t)/\sigma(gg/tt \to h^0_{\rm SM})$ 
as a function of the produced particle mass $m=(m_{A^0_t}, m_{h_{_{\rm SM}}})$.
The mass range analyzed here has been restricted to $563\,\GeV \le m \le 2000\,\GeV$, 
which is indicated from the indirect mass limits in Eqs.(\ref{constraint-mAt}) and (\ref{constraint-mAt:2}). 

From Fig.~\ref{cross-section}, one can see that
the t$\bar{\text{t}}$A production is suppressed compared to the SM Higgs case   
for the mass range constrained by the indirect limits in Eqs.(\ref{constraint-mAt}) and (\ref{constraint-mAt:2}). 
The ggF production is somewhat suppressed compared to the SM Higgs one, 
which is due to the numerical suppression by the overall coupling, 
$g_{A^0_t gg} \propto 2(2)^2 \tan^2\theta$  (in the heavy quark limit) 
when we take $\cos\theta \sim 1$ to be consistent with the indirect limits. 
Since the ggF production is  
much larger than the t$\bar{\text{t}}$H production in the case of the SM Higgs boson, 
the ggF production of the $A^0_t$ is highly dominant enough to neglect 
the t$\bar{\text{t}}$A production at the LHC.

%

\subsection{Current LHC limits on high-mass $A^0_t$}

\begin{figure}[htb]
\begin{center}
\begin{tabular}{cc}
{
\begin{minipage}[t]{0.4\textwidth}
\includegraphics[scale=0.6]{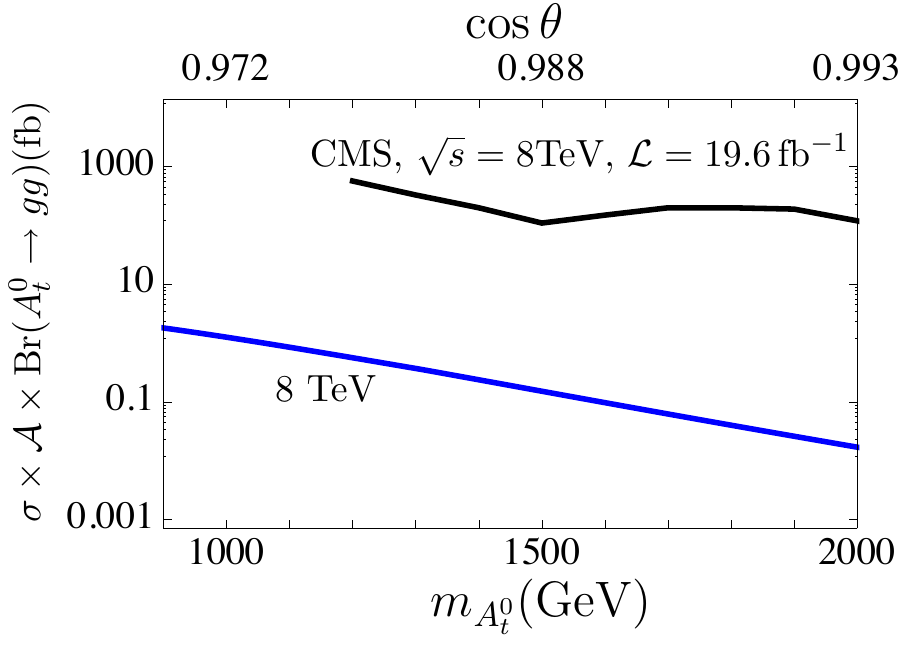} 
\end{minipage}
}
&
{
\begin{minipage}[t]{0.4\textwidth}
\includegraphics[scale=0.6]{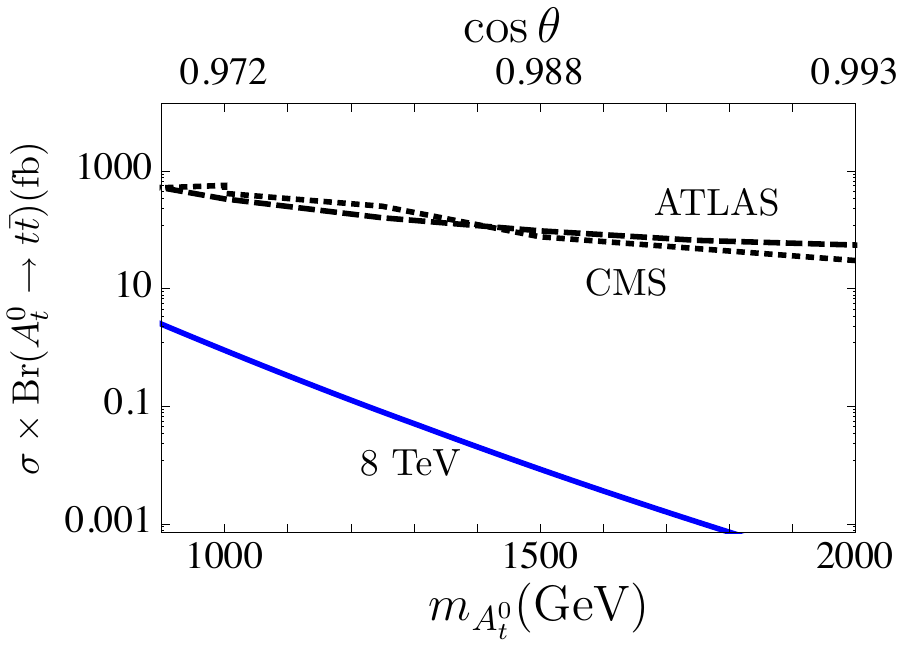} 
\end{minipage}
}
\end{tabular}
\caption[]{
The cross sections, 
$\sigma (gg \to A^0_t) \times \text{Br}(A^0_t \to gg)$ (left panel) 
and 
$\sigma (gg \to A^0_t) \times \text{Br}(A^0_t \to t\bar{t})$ (right panel) 
as a function of $m_{A^0_t}$ in units of fb. 
In the left panel, the observed $95\%\cl$ upper limit (black solid curve) is quoted from 
data on searches for new resonances in dijet mass distribution of gluon-gluon type   
by the CMS experiments 
at $\sqrt{s}=8\,\TeV$ with ${\cal L}=19.6\,{\text{fb}}^{-1}$~\cite{CMS:kxa}, 
where ${\cal A}$ denotes the acceptance, ${\cal A} =0.6$, which is read off from the reference. 
In the right panel, the observed $95\%\cl$ upper limit 
from searches for $Z'$ resonance with $\Gamma_{Z'}/M_{Z'}=1.2\%$ 
by the ATLAS experiments at  
$\sqrt{s}=8\,\TeV$ with ${\cal L}=14\,\text{fb}^{-1}$ data~\cite{TheATLAScollaboration:2013kha}
(black dashed curve) and 
the CMS experiments at 
$\sqrt{s}=8\,\TeV$ with ${\cal L}=19.7\,\text{fb}^{-1}$ data~\cite{Chatrchyan:2013lca} 
(black dotted curve) are also shown. 
\label{cross-section-gg-tt-Zh}}
\end{center}
\end{figure}

We shall discuss the current LHC limits on the high-mass $A^0_t$ 
($1\,\TeV \leq m_{A^0_t} \leq 2\,\TeV$) 
in comparison with the available data on searches for new resonances at LHC Run-I. 
We focus on the $A^0_t$ signals produced via the ggF decaying to 
$t\bar{t}$ and $gg$. 
Figure~\ref{cross-section-gg-tt-Zh} shows the plots of the production cross section 
times branching ratio $\sigma \times \text{Br}(A^0_t \to gg/ t\bar{t})$, 
as a function of $m_{A^0_t}$ in units of fb. 
In the figure the observed $95 \%\cl$ upper limits for each channel have also been plotted, 
which are quoted from 
Refs.\cite{TheATLAScollaboration:2013kha, Chatrchyan:2013lca,CMS:kxa}. 
More on details of the comparison with those data have been given in the caption 
of Fig.~\ref{cross-section-gg-tt-Zh}. 
In computing the ggF production cross section $\sigma(gg \to A^0_t)$, 
we have used the CTEQ6M~\cite{Pumplin:2002vw} for the parton distribution function.
Here we have taken into account the K factor for the ggF production  
of CP-odd scalars with the mass $m_A$,   
$ K^g_A = 1 +(69/4) \alpha_s(m_A)/\pi $~\cite{Djouadi:2005gi}. 
Figure~\ref{cross-section-gg-tt-Zh} implies that 
the high-mass $A^0_t$ has not severely been constrained yet by the LHC Run-I data. 

It is anticipated that the upcoming LHC Run-II will provide 
more stringent constraints or a hint for the discovery of $A^0_t$. 
In particular, 
the searches for CP-odd scalars decaying to $t \bar{t}$ and $gg$ would 
be interesting and challenging to probe the high-mass $A^0_t$ around a few TeV, 
which has not so far been performed. 
The characteristic feature of $A^0_t$ would be seen 
as ``a quite narrow resonance" with $\Gamma_{\rm tot} \ll m_{A^0_t}$ 
in the $t\bar{t}$ and $gg$ mass distributions, 
as is indicated from the left panel of Fig.~\ref{decay-At0}.   

\subsection{Current LHC limits on low-mass $A^0_t$}

We next discuss the LHC discovery channel of the low-mass 
$A^0_t$ ($563\,\GeV \leq m_{A^0_t} \leq 1\,\TeV$). 
In the case of the low-mass $A^0_t$, 
one can see from Fig.~\ref{decay-At0} that 
an interesting channel is the $A^0_t \to Z h^0_t$ 
having the branching ratio $\text{Br}(A^0_t \to Zh^0_t) =20-40\,\%$, 
which would be large enough to be accessible at the LHC. 

\begin{figure}[htbp]
\begin{center}
\begin{tabular}{cc}
{
\begin{minipage}[t]{0.4\textwidth}
\includegraphics[scale=0.6]{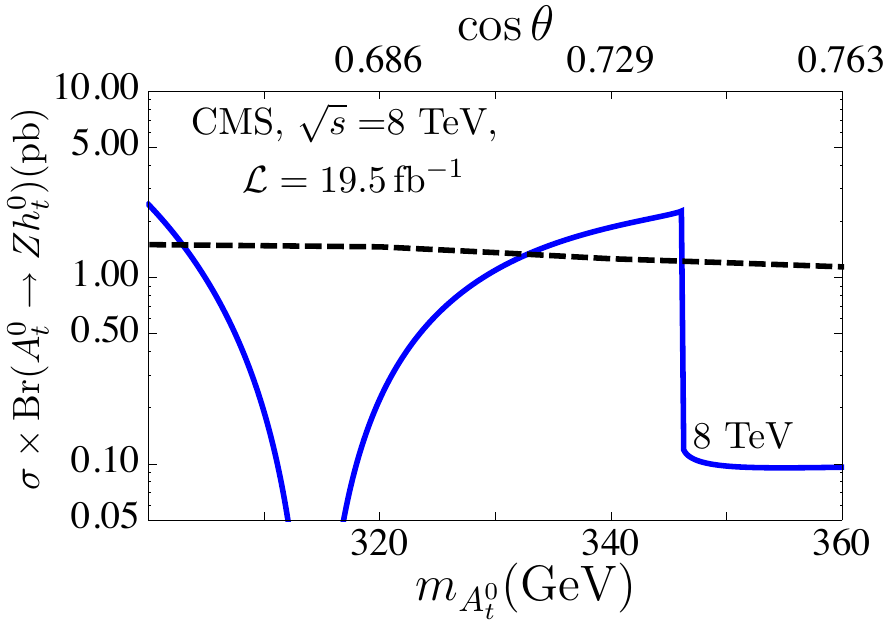} 
\end{minipage}
}
&
{
\begin{minipage}[t]{0.4\textwidth}
\includegraphics[scale=0.6]{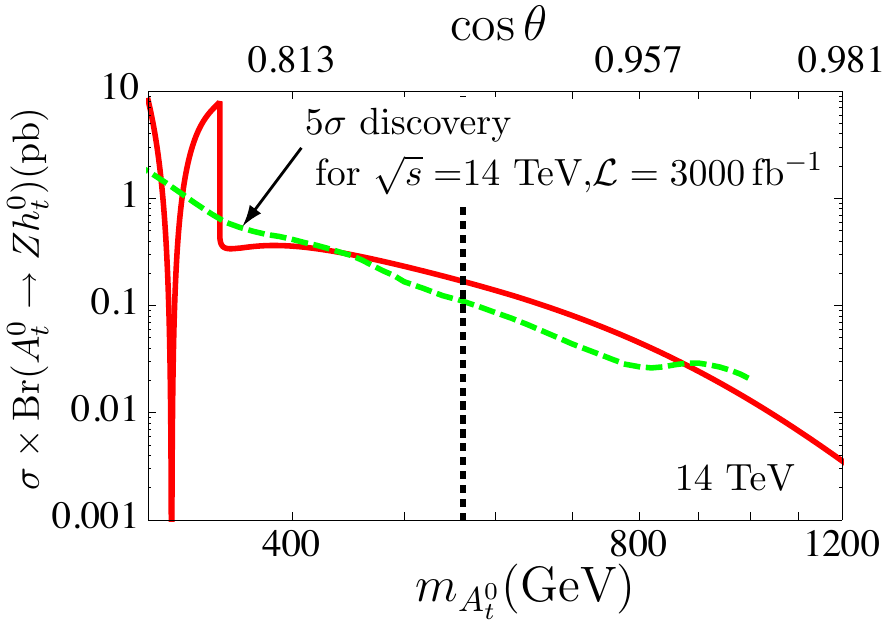} 
\end{minipage}
}
\end{tabular}
\caption[]{
Left panel:
The plot of $\sigma_{\rm ggF} (pp \to A^0_t) \times \text{Br}(A^0_t \to Zh^0_t)$ 
as a function of $m_{A^0_t}$ in units of pb 
for a low-mass range $m_{A^0_t} (m_{A^0_t} \leq 360\,\GeV)$.
The observed $95\%\cl$ upper limit from data on searches for extended Higgs sectors 
by the CMS experiments 
at $\sqrt{s}=8\,\TeV$ with ${\cal L}=19.5\,\text{fb}^{-1}$~\cite{CMS:2013eua} 
has also been shown by the black dashed curve. 
Right panel: 
The expected signal strength of 
$\sigma_{\rm ggF} (pp \to A^0_t) \times \text{Br}(A^0_t \to Zh^0_t)$ 
at the 14 TeV LHC 
for a mass range $m_{A^0_t} \leq 1\,\TeV$ (red solid curve).  
Also,
the curve corresponding to the $5\sigma$ discovery at ${\cal L}=3000\,{\text{fb}}^{-1}$ 
provided by the CMS simulation \cite{CMS:2013dga} 
(green dashed curve) has been displayed.
The vertical dotted line corresponds to 
the indirect mass limit in Eq.(\ref{constraint-mAt:2}).
\label{cross-section-AZh}
}
\end{center}
\end{figure}

In the left panel of Fig.~\ref{cross-section-AZh}, 
we make a plot of the production cross section times the branching ratio of $A^0_t$ 
for the $Zh^0_t$ channel,  
$\sigma \times \text{Br}(A^0_t \to Zh^0_t)$, 
as a function of $m_{A^0_t}$ in units of pb, 
together with the observed limit from 
the currently available data on  
searches for extended Higgs sectors by the CMS experiments
at $\sqrt{s}=8\,\TeV$ with ${\cal L}=19.5\,\text{fb}^{-1}$~\cite{CMS:2013eua}. 
Here we have allowed a light $A^0_t$ having the mass slightly 
off from the indirect limit in Eq.(\ref{constraint-mAt:2}), 
since the current LHC bound has not reached a higher-mass region 
such as $m_A \geq 563\,\GeV$. 
The current limit requires the low-mass $A^0_t$ to have  
\beq 
303\,\GeV \leq m_{A^0_t} \leq 333\,\GeV
\quad \text{or} \quad
2m_t \simeq 346\,\GeV \leq m_{A^0_t}
\,,\label{direct-constraint-At0}
\eeq
in which the latter case is consistent with the indirect limit in Eq.(\ref{constraint-mAt:2}).
The small allowed window for a low-mass region 
$(303\,\GeV \leq m_{A^0_t} \leq 333\,\GeV)$ has been present 
because of the dip in Eq.(\ref{decay-At-Zh}) [see Eq.(\ref{dip})].

The $A^0_t$ branching ratio for the $Z h^0_t$ mode is comparable with 
the branching ratio of the CP-odd Higgs boson ($A^0$) decaying to $Z h^0$ in the MSSM/2HDM 
with $\tan \beta =3$~\cite{Djouadi:2005gi},
so it is interesting to compare these signals at the same mass.  
It turns out that 
the production cross section of $A^0_t$ is actually larger than that of $A^0_t$ 
in the MSSM/2HDM for the low-mass region; 
indeed, 
the ratio of the ggF production cross sections of 
$A^0_t$ to that of $A^0$ evaluated at the same mass goes like 
\beq
\frac{\sigma(gg \to A^0_t)}{\sigma(gg \to A^0)}
\simeq 
\frac{2\tan^2\theta}{\cot^2\beta}
\,,\label{gg-ratio-TMP-MSSM}
\eeq
where the heavy quark mass limit $m_{t} \to \infty$ has been taken. 
Taking $m_{A^0_t} = 600\,\GeV$ as a sample benchmark point for the low-mass $A^0_t$, 
we may numerically estimate the ratio in Eq.(\ref{gg-ratio-TMP-MSSM}) to get 
\beq
\frac{\sigma(gg \to A^0_t)}{\sigma(gg \to A^0)}
\simeq 3.2 
\,, 
\qquad \textrm{for} \qquad \tan \beta =3 
\,.\label{enhancement:ggF}
\eeq
This implies that 
the search for $A^0_t$ would be more accessible than $A^0$ through the $Z h^0$ channel; 
the discovery of $A^0_t$ would be possible earlier than that of $A^0$.  

In light of the LHC experiment with the high statistics, 
in the right panel of Fig.~\ref{cross-section-AZh}, we plot   
$\sigma_{\rm ggF} \times \text{Br}(A^0_t \to Zh^0_t)$ 
as a function of $m_{A^0_t}$ in units of pb at $\sqrt{s}=14$ TeV (red solid curve).  
Also, the $5 \sigma$ discovery potential (green dashed curve)
for pseudo scalars in the $Zh$ channel at ${\cal L}=3000\,{\text{fb}}^{-1}$ 
provided from the CMS simulation~\cite{CMS:2013dga} has been shown. 
From the right panel of Fig.~\ref{cross-section-AZh}, 
we see that the low-mass $A^0_t$ with the mass in a range
\beq 
563\,\GeV \leq m_{A^0_t} \leq 875\,\GeV
\eeq
is expected to be discovered at the upcoming LHC experiments.

\section{Summary}
\label{summary}

In summary, 
we discussed the LHC phenomenologies of the Top-Mode Pseudos,
$h^0_t$ (CP-even scalar, tHiggs), and $A^0_t$ (CP-odd scalar),
arising as composite PNGBs in the model recently proposed 
in a framework of the top quark condensation. 
We first analyzed the tHiggs $h^0_t$ couplings to the SM particles   
to compare them with the currently available data on 
the Higgs coupling measurements at the LHC. 
It was shown that the tHiggs can be consistent with the 126 GeV Higgs boson, 
allowing the amount of deviation controlled 
by a model parameter $\cos\theta \geq 0.91$ at $95\,\%\cl$
[Eq.(\ref{constraint-cos:2})]. 
The bound on $\cos\theta$ was converted into an indirect limit on 
the mass of the CP-odd Top-Mode Pseudo $A^0_t$ 
through the Top-Mode Pseudo mass formula [Eq.(\ref{mass-ht0})], 
leading to the lower bound $m_{A^0_t} \geq 563 \,\GeV$ [Eq.(\ref{constraint-mAt:2})]. 

We explored the direct searches for $A^0_t$ at the LHC 
by explicitly calculating the relevant production cross sections and partial decay widths. 
The total width of $A^0_t$ was shown to be quite smaller than that of the SM Higgs boson 
and 
other CP-odd scalars like $A^0$ as in the MSSM/2HDM (left panel of Fig.~\ref{decay-At0}). 
This feature is still operative even for a high-mass region $m_{A^0_t} > 1\,\TeV$. 
This is essentially due to the intrinsic feature of the Top-Mode Pseudos 
characterized by the mass formula [Eq.(\ref{mass-ht0})],  
which allows us to express the overall coupling of $A^0_t$, $\sin\theta$, 
in terms of $1/m_{A^0_t}$, 
leading to the significant suppression of the total width in the high-mass region: 
In that sense, the weak coupling nature of $A^0_t$ is ensured by the mass formula. 
The branching fraction of $A^0_t$ was discussed by dividing into 
two cases (right panel of Fig.~\ref{decay-At0}): 
i) high-mass $A^0_t$ with $m_{A^0_t} \geq 1\,\TeV$, 
where 
$A^0_t$ decays to the digluon ($\sim 63\,\%$) 
and 
$A^0_t$ decays to the tHiggs associated with the $Z$ boson ($\sim 16\%$) 
for $m_{A^0_t} \simeq 1.2\,\TeV$, and
ii) low-mass $A^0_t$ with the mass in the range $563\,\GeV \leq m_{A^0_t} \leq 1\,\TeV$, 
where $A^0_t$ mainly decays to $t\bar{t}$ ($\sim 66\%$) 
and the tHiggs associated with the $Z$ boson ($\sim 27\%$) 
for $m_{A^0_t} = 600\,\GeV$.
It was also found that the LHC production of $A^0_t$ is highly dominated 
by the ggF process for both the low-mass 
and 
high-mass cases (Fig.~\ref{cross-section}). 

We then placed the current LHC limit on $m_{A^0_t}$ 
by using the currently available data on searches for new resonances in several channels 
(Fig.~\ref{cross-section-gg-tt-Zh} and the left panel of Fig.~\ref{cross-section-AZh}) 
to find that all the direct limits are weaker than the indirect limit 
from the Higgs coupling measurements [Eq.(\ref{constraint-mAt:2})].  

In light of the upcoming LHC experiment with higher statistics, 
the searches for CP-odd scalars decaying to $t \bar{t}/gg$ would 
be interesting and challenging to probe the $A^0_t$ 
with the mass $m_{A^0_t} \geq 1\,\TeV$, 
which has not so far been performed. 
The characteristic feature of $A^0_t$ would be seen as 
a quite narrow resonance 
with $\Gamma_{\rm tot} \simeq 0.1\,\GeV$ 
in the $t\bar{t}/gg$ mass distribution (left panel of Fig.~\ref{decay-At0}).    
Or a somewhat light $A^0_t$ with the mass 
in a range of $563 \,\GeV \leq m_{A^0_t} \leq 1\,\TeV$ 
is expected to be observed as a quite narrow resonance 
in the channel decaying to $Z h^0$ produced from the ggF process.  
Such a light $A^0_t$ could be observed earlier than 
other CP-odd scalars such as $A^0$ in the MSSM/2HDM, 
due to the larger ggF production cross section [Eq.(\ref{enhancement:ggF})]. 
We examined the discovery potential of 
the low-mass $A^0_t$ decaying into the $Zh^0_t$ 
at the 14 TeV LHC 
(right panel of Fig.~\ref{cross-section-AZh}). 
A light $A^0_t$ with the mass in a range $563\,\GeV \leq m_{A^0_t} \leq 875\,\GeV$ 
can be discovered at the $5 \sigma$ level with  
${\cal L}=3000\,{\text{fb}}^{-1}$. 

More precise estimates on the $A^0_t$ discovery potential at LHC Run-II 
will be pursued in another publication.

Throughout the present paper, 
we have employed the nonlinear sigma model 
by integrating out the heavy Higgs boson $H^0_t$ at around ${\cal O}(1)\,\TeV$. 
One could also study the LHC phenomenology of the $H^0_t$ based on the linear sigma model, 
instead of the nonlinear realization. 
If one examines Appendix A of  Ref.\cite{Fukano:2013aea}, 
one would notice that the $H^0_t$ with the mass of coupling property for the 
SM particles can be found just by rotating the tHiggs couplings 
by the angle $\theta$, namely, 
replacing the overall angle $\cos\theta$ by $\sin\theta$. 
Taking into account the experimental constraint $\cos\theta \sim 1$, 
one would then find that all the production cross sections regarding the $H^0_t$ 
are suppressed by the overall factor of $\sin^2\theta$, 
compared to the SM-like Higgs case including the tHiggs. 
More precise arguments on this topic can be done in a way 
similar to the analyses on the top-Higgs boson as done in Refs.~%
\cite{Chivukula:2000px,Chivukula:2011dg,Chivukula:2012cp}, 
which deserves another publication.

\section*{Acknowledments}
This work was supported in part by the JSPS Grant-in-Aid for Scientific Research (S), Grant No. 22224003. 

\bibliography{ref-SMHiggs,ref-EWPT,ref-2HDM,ref-TC,ref-TMSM,ref-PDG-PDF,ref-Higgs-ATLAS,ref-Higgs-CMS,ref-LHC-woSMHiggs}
\end{document}